\documentstyle[aps, prb, tighten]{revtex}
\begin{document}
\draft
\hoffset= -2.5mm

\title{
High-field noise in degenerate and mesoscopic systems \\}

\author{F. Green\cite{fgemail} \\}
\address{
GaAs IC Prototyping Facility Program,
CSIRO Telecommunications and Industrial Physics,\\
PO Box 76, Epping NSW 1710, Australia \\}

\author{M. P. Das\cite{mpdemail}  \\}
\address{
Department of Theoretical Physics,
Research School of Physical Sciences and Engineering, \\
The Australian National University,
Canberra ACT 0200, Australia \\}

\maketitle

\begin{abstract}
We analyse high-field current fluctuations in metallic systems
by direct mapping of the Fermi-liquid correlations to
the semiclassical nonequilibrium state. We give three applications.
First, for bulk conductors, we show that there is
a unique nonequilibrium analogue to the fluctuation-dissipation theorem
for thermal noise. With it, we calculate suppression of
the excess hot-electron term by Pauli exclusion.
Second, in the degenerate mesoscopic regime,
we argue that shot noise and thermal noise are incommensurate.
They cannot be connected by a smooth, universal interpolation formula.
This follows  from their contrasting responses to Coulomb screening.
We propose an experiment to test this mismatch.
Third, we carry out an exact model calculation of high-field shot noise
in narrow mesoscopic wires. We show that a distinctive
mode of suppression arises from the structure of the Boltzmann
equation in two and three dimensions.
In one dimension such a mode does not exist.

\end{abstract}


\section{Introduction}

High-field noise in degenerate conductors still lacks a systematic
theoretical description, despite its importance for microelectronics.
\cite{mc}
Here we advance a practicable theory of nonequilibrium
fluctuations in metallic systems, accounting for the dominant
Fermi-liquid behaviour of the electrons.
\cite{nozpin}
There is no electronic property of a metal near
equilibrium that is not governed by Pauli exclusion,
from the microscopic level to the bulk.
\cite{ashmer}
In this paper we analyse how degeneracy determines nonequilibrium
current noise over a wide range of length scales,
even at high fields.

Technological developments of late have led to a variety of
delicate measurements of transport and noise in many different
mesostructures.
\cite{rez,ksgje,smd,sbkpr,poth,depic,sami}
Alongside the experiments
there has been much theoretical activity.
\cite{theor,lesovik,but,beebut,thmldr,nagaev1,djb2}
A particular topic, still being elaborated,
is the behaviour of current fluctuations in mesoscopic conductors.
These are typically shorter than the inelastic mean free path
but longer than that for elastic processes. They are in
the regime of diffusive transport, where
two diverse understandings predominate.
One technique is the quantum-transmission (Landauer) method
\cite{ldr57}
as applied to fluctuations by
Lesovik,
\cite{lesovik}
Beenakker and B\"uttiker,
\cite{but,beebut}
Martin and Landauer,
\cite{thmldr}
and many others.
\cite{theor}
The second approach
is the semiclassical-transport (Boltzmann) method
\cite{kogan}
associated with
Nagaev
\cite{nagaev1}
and de Jong and Beenakker.
\cite{theor,djb2}
Both view a mesoscopic wire as an assembly of
individual elastic scatterers in a bath of free carriers.
The formalisms, however, are quite dissimilar.
\cite{shimi}

In the Landauer model, multiple scattering preserves the coherence
of single-particle propagation; the only way in which
current fluctuations can lose correlation strength is
by interplay of the transmission amplitudes and
nonlocal Pauli blocking between state occupancies in
the source and drain leads.
\cite{landauer}
In the Boltzmann model, incoherence enters from the start
through the Stosszahlansatz. To study the fluctuations,
the stochastic collision term is supplemented with
a set of Langevin flux sources
whose phenomenology is that of classical shot noise;
\cite{kogan}
their self-correlations reflect the sporadic timing of
elementary encounters between discrete wave packets and scatterers.
The induced fluctuations lose correlation strength
when diffusive elastic scattering is locally modified by Pauli exclusion.
\cite{theor,yam,landauer2}

Regardless of their differences, both methods can
explain observations such as the threefold suppression
of shot noise when elastic scattering prevails.
\cite{smd}
Because the phase-coherent model is fully quantum mechanical,
its fluctuation structure is natural, not imposed, and
its predictions enjoy a definitive status.
On the other hand, semiclassical phenomenology is the
more natural tool when inelastic (hence irreducibly
phase-breaking) collisions are important, as in high-field transport.

Our paper investigates fluctuations beyond the elastic weak-field limit.
A mesoscopic sample is easily driven into the high-field regime;
some tens of millivolts across a length of 100 nm will do it.
\cite{ferry}
Nevertheless, although nonequilibrium noise is
a unique source of dynamical information
out of reach to linear-response theory,
\cite{sw2}
none of the existing models has been pushed
substantially beyond its low-field perturbative regime.
The Landauer and Boltzmann-Langevin formalisms
each suffers from its own obstacles to addressing
strongly nonequilibrium effects, as does the quantum-kinetic
theory of Altshuler, Levitov, and Yakovets,
\cite{aly}
which tries to unify them.
There is, therefore, a real need for another approach:
one that is nonperturbative in the driving field.
This need has been reasserted very recently by numerical evidence
of a rich structure for shot noise at high fields,
even in nondegenerate systems.
\cite{reggi2}

We propose a direct and versatile method
for treating noise semiclassically from equilibrium up to high fields,
building on Fermi-liquid theory
\cite{nozpin,rickayzen}
and a family of Green functions for the linearised transport equation.
Such functions have been studied by Kogan and Shul'man,
\cite{kogan}
Gantsevich, Gurevich, and Katilius,
\cite{ggk}
and by Stanton and Wilkins, in great detail, for semiconductors.
\cite{sw2,sw0}
Unlike Boltzmann-Langevin, this approach is
not limited to the specifically Boltzmannian form
of the collision integral.
\cite{sw0}
For every collisional approximation that can be used
in the one-particle transport equation,
there is a systematic construction for the
two-particle Boltzmann-Green functions (BGFs).
These then generate both the steady-state and transient
nonequilibrium fluctuations, respecting the conservation laws
as well as the one-body collisional structure
and bringing great flexibility to noise calculations.
An application to the noise perfomance of heterojunction
transistors is given by Green and Chivers.\cite{gc}

In Section II we present a general framework for noise
in small metallic systems, including Coulomb effects
\cite{reggi2,nvhcoul,rggcoul,ngvcoul}
where we identify two distinct mechanisms for screening.
Sec. III contains the major applications of our theory.
First, we extend the fluctuation-dissipation relation
for thermally driven noise to nonequilibrium conductors,
\cite{vvt,nougier,nerlul}
highlighting the role of Pauli exclusion in suppressing
hot-electron effects in the bulk noise spectrum.
\cite{mc}
We then discuss the many-body origin of shot noise and
argue that, in their sharply contrasting responses to Coulomb screening,
shot noise and thermal noise display quite different physics.
We propose a simple experimental test of this difference.
In Sec. IV we make an exact computation of shot noise
within the Drude picture of a conducting wire. We include the effects
of finite wire thickness on carrier motion,
a significant source of shot-noise suppression 
at high currents that is unrelated to diffusive elastic scattering
and to Coulomb screening.
We sum up in Sec. V.

\section{Theory}

The theoretical discussion is in four parts. We begin
by formulating the transport problem as a direct mapping of the
electron Fermi liquid to its nonequilibrium steady state.
Second, we describe the steady-state fluctuations.
Third, we discuss the dynamic fluctuations and their formal
connection with the steady state. Last, we incorporate
Coulomb screening into the nonequilibrium structure.
Our end product is a complete expression for the current
autocorrelation, which determines the noise.

\subsection{Transport Model}

The semiclassical Boltzmann transport
equation (BTE) for the electron distribution function
$f_{\alpha} (t) \equiv f_{s}({\bf r}, {\bf k}, t)$ is

\begin{eqnarray}
{\left[
{ {\partial}\over {\partial t} } +
  { {\bf v}_{{\bf k} s } }{\bbox \cdot}
 { {\partial}\over {\partial {\bf r} }} -
  {
   { { e{\bf E}({\bf r},t) }\over {\hbar} }{\bbox \cdot}
  }
   { {\partial}\over {\partial {\bf k}} }
\right]} f_{\alpha}(t)
=
-\sum_{\alpha'}
{\Bigl[
W_{\alpha' \alpha} (1 - f_{\alpha'}) f_{\alpha}
- W_{\alpha \alpha'} (1 - f_{\alpha}) f_{\alpha'}
\Bigr]}.
\label{AX1}
\end{eqnarray}

\noindent
Label $\alpha = \{{\bf r}, {\bf k}, s\}$ denotes a
point in single-particle phase space, while sub-label {\em s}
indexes both the discrete subbands (or valleys) of
a multi-level system and the spin state.
The system is acted upon by the total
internal field ${\bf E}({\bf r},t)$.
We study single-particle scattering, with a rate
$W_{\alpha \alpha'} \equiv
\delta({\bf r} - {\bf r'})
W_{s s'}({\bf r}, {\bf k}, {\bf k'})$ that is local in real space,
independent of the driving field, and that satisfies detailed balance:
$W_{\alpha' \alpha} (1 - f^{\rm eq}_{\alpha'}) f^{\rm eq}_{\alpha}
= W_{\alpha \alpha'} (1 - f^{\rm eq}_{\alpha}) f^{\rm eq}_{\alpha'}$
where $f^{\rm eq}_{\alpha}$ is the equilibrium distribution.
In a system with $\nu$ dimensions, we make
the following correspondence for the identity operator:

\[
\delta_{\alpha \alpha'}
\equiv \delta_{s  s'}
{\left\{ { {\delta_{{\bf r} {\bf r'}}}\over
	   {\Omega({\bf r})} } \right\}}
{\left\{ \Omega({\bf r}) \delta_{{\bf k} {\bf k'}} \right\}}
\longleftrightarrow \delta_{s  s'}
\delta({\bf r} - {\bf r'})
(2\pi)^{\nu} \delta({\bf k} - {\bf k'}).
\]

\noindent
The volume $\Omega({\bf r})$
of a local cell in real space becomes the measure for
spatial integration, while its inverse defines the
scaling in wave-vector space
for the local bands $\{ {\bf k}, s \}$.

The first step is to construct the steady-state solution
$f_{\alpha} \equiv f_{\alpha}(t \to \infty)$ explicitly
from $f^{\rm eq}$, which satisfies the equilibrium, collisionless
form of Eq. (\ref{AX1}):

\begin{equation}
  { {\bf v}_{{\bf k} s } }{\bbox \cdot} 
 {{\partial f^{\rm eq}_{\alpha} }\over { \partial {\bf r} }} -
  {
   { { e{\bf E}_0({\bf r}) }\over {\hbar} }{\bbox \cdot}
  }
   { {\partial f^{\rm eq}_{\alpha} }\over {\partial {\bf k}} }
= 0.
\label{AX1.1}
\end{equation}

\noindent
The internal field ${\bf E}_0({\bf r})$ is defined in the absence of a
driving field. The quantities $f^{\rm eq}$ and ${\bf E}_0$
are linked self-consistently by the usual constitutive relations,
the first being the Poisson equation

\begin{equation}
{\partial\over {\partial {\bf r}} } {\bbox \cdot}
\epsilon {\bf E}_0 = -4\pi e
{\Bigl(  {\langle f^{\rm eq}({\bf r}) \rangle} - n^+({\bf r}) \Bigr)}
\label{poissoneq}
\end{equation}

\noindent
in terms of the dielectric constant $\epsilon({\bf r})$,
the electron density
$\langle f^{\rm eq}({\bf r}) \rangle \equiv
{\Omega({\bf r})}^{-1}{\sum}_{{\bf k},s}
f^{\rm eq}_{\alpha}$,
and the positive background density $n^+({\bf r})$,
which will remain unchanged throughout our calculations.
\cite{embed}
Normalisation to the total particle number is
$\sum_{\bf r} \Omega({\bf r}) {\langle f^{\rm eq}({\bf r}) \rangle} = N$.
The second relation is the form of the equilibrium function itself,

\begin{equation}
f^{\rm eq}_{\alpha} ~=~
{\left[
1 + \exp \!
 {\left(
   { {\varepsilon_{\alpha} - \phi_{\alpha}}\over k_{B}T }
  \right)}
\right]}^{-1},
\label{AX1.2}
\end{equation}

\noindent
in which the local conduction-band energy
$\varepsilon_{\alpha} = \varepsilon_s({\bf k}; {\bf r})$
can have band parameters that depend on position.
The locally defined Fermi level
$\phi_{\alpha} = \mu - V_0({\bf r})$ is the difference
of the global chemical potential $\mu$ and the
electrostatic potential $V_0({\bf r})$,
whose gradient is $e{\bf E}_0({\bf r})$.

Define the difference function 
$g_{\alpha} = f_{\alpha} - f^{\rm eq}_{\alpha}$.
From each side of Eq. (\ref{AX1}) in the steady state,
subtract its equilibrium counterpart. We obtain

\begin{eqnarray}
  { {\bf v}_{{\bf k} s } }{\bbox \cdot} 
 {{\partial g_{\alpha} }\over { \partial {\bf r} }} -
  {
   { { e{\bf E}({\bf r}) }\over {\hbar} }{\bbox \cdot}
  }
   { {\partial g_{\alpha} }\over {\partial {\bf k}} } =&&
  {
   { { e({\bf E} - {\bf E}_0) }\over {\hbar} }{\bbox \cdot}
  }
   { {\partial f^{\rm eq}_{\alpha}
     }\over {\partial {\bf k}} }
- \sum_{\alpha'}
( W_{\alpha' \alpha} g_{\alpha} - W_{\alpha \alpha'} g_{\alpha'} )
\cr
&&+ \sum_{\alpha'}
( W_{\alpha' \alpha} - W_{\alpha \alpha'} )
( f^{\rm eq}_{\alpha'} g_{\alpha}
+ g_{\alpha'} f^{\rm eq}_{\alpha} + g_{\alpha} g_{\alpha'} ).
\label{AX3}
\end{eqnarray}

\noindent
The solutions to Eqs. (\ref{AX1.1}) and
(\ref{AX3}) are determined by the
asymptotic conditions in the source and drain reservoirs,
be it at equilibrium or with an external electromotive force.
Our active region includes the carriers within the
source and drain terminals out to several
screening lengths, so that local fields
are negligible at the interfaces with the reservoirs.
In practice we assume that all fields are shorted out so that
${\bf E}({\bf r}) = {\bf E}_0({\bf r}) = {\bf 0}$
beyond the boundaries. Gauss's theorem implies that this
bounded system remains neutral overall:

\begin{equation}
\sum_{\bf r} \Omega({\bf r})\langle g ({\bf r}) \rangle
= \sum_{\alpha} g_{\alpha} = 0.
\label{gauss}
\end{equation}

We put Eq. (\ref{AX3}) into integro-differential form, with
an inhomogeneous term explicitly dependent on $f^{\rm eq}$:

\begin{equation}
\sum_{\alpha'}
B[W^A f]_{\alpha \alpha'} g_{\alpha'}
= { { e{\bf {\widetilde E}} ({\bf r})}
\over {\hbar} }{\bbox \cdot}
   { {\partial  f^{\rm eq}_{\alpha}
}\over {\partial {\bf k}} }
+ \sum_{\alpha'}
W^A_{\alpha \alpha'} g_{\alpha'} g_{\alpha}
\label{AX5}
\end{equation}

\noindent
where
${\bf {\widetilde E}}({\bf r}) = {\bf E}({\bf r}) - {\bf E}_0({\bf r})$
is the local field induced by the external electromotive potential
and $B[W^A f]$ is the linearised Boltzmann operator

\begin{eqnarray}
B[W^A f]_{\alpha \alpha'}
\buildrel \rm def \over =&&
\delta_{\alpha \alpha'}
{\left[
{ {\bf v}_{{\bf k'} s'} }{\bbox \cdot} 
 {{\partial }\over { \partial {\bf r'} }} -
  {
   { { e{\bf E}({\bf r'}) }\over {\hbar} }{\bbox \cdot}
  }
   { {\partial }\over {\partial {\bf k'}} } +
  \sum_{\beta} ( W_{\beta \alpha'}
- W^A_{\beta \alpha'} f_{\beta} )
\right]}
\cr
&&
- W_{\alpha \alpha'} + W^A_{\alpha \alpha'} f_{\alpha},
\label{bop}
\end{eqnarray}

\noindent
and  $W^A_{\alpha \alpha'} = W_{\alpha \alpha'} - W_{\alpha' \alpha}$.
Note that $W^A = 0$ if the scattering is elastic.
To represent the physical solution,
$g$ must vanish with ${\widetilde E}$ in the equilibrium limit.
This is guaranteed by the Poisson equation

\begin{equation}
{\partial\over {\partial {\bf r}} } {\bbox \cdot}
\epsilon {\bf {\widetilde E}}
= -4\pi e
{\Bigl( \langle f({\bf r}) \rangle
- \langle f^{\rm eq}({\bf r}) \rangle \Bigr)}
= -4\pi e \langle g ({\bf r}) \rangle.
\label{poiss}
\end{equation}

\subsection{Boltzmann-Green Functions}

To calculate the adiabatic response of the system
about its nonequilibrium steady state
we introduce the Boltzmann-Green function
\cite{fg1}

\begin{equation}
G_{\alpha \alpha'} \buildrel \rm def \over =
{ { \delta g_{\alpha} }\over
  { \delta f^{\rm eq}_{\alpha'} } },
\label{AX6.1}
\end{equation}

\noindent
with a global constraint following
directly from Eq. (\ref{gauss}):

\begin{equation}
\sum_{\alpha} G_{\alpha \alpha'} = 0.
\label{dgauss}
\end{equation}

\noindent
The equation of motion for $G$ is derived by taking variations
on both sides of Eq. (\ref{AX3}):

\begin{equation}
\sum_{\beta}
B[W^A f]_{\alpha \beta} G_{\beta \alpha'}
= \delta_{\alpha \alpha'}
{\left[
{ { e{\bf {\widetilde E}} ({\bf r'})}
\over {\hbar} }{\bbox \cdot}
   { {\partial }\over {\partial {\bf k'}} }
+ \sum_{\beta} W^A_{\beta \alpha'} g_{\beta}
\right]} - W^A_{\alpha \alpha'} g_{\alpha}.
\label{AXG}
\end{equation}

\noindent
The variation is restricted by excluding the reaction of the local fields
${\bf E}_0({\bf r})$ and ${\widetilde {\bf E}}({\bf r})$.
This means that $G$ is a response function free of Coulomb screening.
Here we treat the electrons
as an effectively neutral Fermi liquid; in Section IID we
give the complete fluctuation structure, with Coulomb effects.

All of the steady-state fluctuation properties
induced by the thermal background
are specified in terms of $G$
and the equilibrium fluctuation $\Delta f^{\rm eq}$.
The equilibrium fluctuation
is the proper particle-particle correlation
in the static long-wavelength limit,
normalised by the thermal energy $k_B T$. In the Lindhard
approximation
\cite{nozpin}
this is

\begin{equation}
\Delta f^{\rm eq}_{\alpha} \equiv
k_B T { {\partial f^{\rm eq}_{\alpha} }
      \over {\partial \phi_{\alpha}} }
= f^{\rm eq}_{\alpha}(1 - f^{\rm eq}_{\alpha}).
\label{AX7.0}
\end{equation}

\noindent
When there are strong exchange-correlation interactions,
this two-body expression is renormalised by a factor
dependent on the Landau quasiparticle parameters.
\cite{nozpin,rickayzen}
In this work we consider free electrons only.

Define the two-particle fluctuation function
$\Delta f^{(2)}_{\alpha \alpha'} \equiv
( \delta_{\alpha \alpha'} + G_{\alpha \alpha'} )
\Delta f^{\rm eq}_{\alpha'}$.
The steady-state distribution of the particle-number fluctuation
is the sum of all of the two-body terms:

\begin{equation}
\Delta f_{\alpha}
= \sum_{\alpha'} \Delta f^{(2)}_{\alpha \alpha'}
= \Delta f^{\rm eq}_{\alpha}
+ \sum_{\alpha'} G_{\alpha \alpha'} \Delta f^{\rm eq}_{\alpha'}.
\label{AX7}
\end{equation}

\noindent
The nonequilibrium fluctuation $\Delta f$, manifestly a
linear functional of its equilibrium Fermi-Dirac form,
is the exact solution to the linearised Boltzmann equation:

\begin{equation}
\sum_{\beta} B[W^A f]_{\alpha \beta} {\Delta f}_{\beta} = 0.
\label{AX7.B}
\end{equation}

\noindent
Charge neutrality implies that the total
fluctuation strength over the sample,
$\Delta N = {\sum}_{\bf r} \Omega({\bf r})
\langle \Delta f ({\bf r}) \rangle$, is conserved.
This constrains both steady-state and dynamical fluctuations.

Calculation of the dynamic response requires
the time-dependent Boltzmann-Green function
\cite{kogan}

\begin{equation}
R_{\alpha \alpha'}(t - t')
\buildrel \rm def \over =
\theta(t - t')
{ {\delta f_{\alpha}(t)}\over
  {\delta f_{\alpha'}(t')} },
\label{AX8}
\end{equation}

\noindent
with initial value
$R_{\alpha \alpha'}(0) = \delta_{\alpha \alpha'}$.
As with $G$, the variation is restricted.
The linearised BTE satisfied by $R(t - t')$
is derived from Eq. (\ref{AX1})
and takes the form

\begin{equation}
\sum_{\beta}
{\left\{
\delta_{\alpha \beta} {{\partial}\over {\partial t}}
+ B[W^A f]_{\alpha \beta}
\right\}} R_{\beta \alpha'}(t - t') =
\delta(t - t') \delta_{\alpha \alpha'}.
\label{drdt}
\end{equation}

\noindent
Summation over $\alpha$ on both sides of this equation leads to
conservation of normalisation:
\cite{kogan}

\begin{equation}
\sum_{\alpha} R_{\alpha \alpha'}(t - t') = \theta(t - t').
\label{rgf1}
\end{equation}

\noindent
The time-dependent BGF is a two-point correlation.
It tracks the history
of a free electron in state $\alpha'$ added to the system
at time $t'$; the probability of finding the electron in
state $\alpha$ at time $t$ is just
$R_{\alpha \alpha'}(t - t')$. In the long-time limit
Eq. (\ref{drdt}) goes to its steady-state form, and so
$R_{\alpha \alpha'}(t \to \infty) \propto \Delta f_{\alpha}$,
the solution to the steady-state linearised equation.
\cite{kogan}
Together with Eq. (\ref{rgf1}) this gives the identity

\begin{equation}
R_{\alpha \alpha'}(t \to \infty)
= { {\Delta f_{\alpha}}\over {\Delta N}}.
\label{rgf2}
\end{equation}

All of the time-dependent fluctuation properties
induced by the thermal background
are specified in terms of $R$
and the steady-state nonequilibrium fluctuation $\Delta f$.
From the dynamical two-particle distribution,
\cite{ggk}
that is $\Delta f_{\alpha \alpha'}^{(2)}(t) \equiv
R_{\alpha \alpha'}(t) \Delta f_{\alpha'}$,
one can construct the lowest-order moment

\begin{equation}
\Delta f_{\alpha}(t)
= \sum_{\alpha'} {\Delta f^{(2)}_{\alpha \alpha'}}(t)
\label{rgf3}
\end{equation}

\noindent
in analogy with Eq. (\ref{AX7}).
Equation (\ref{drdt}) has an adjoint,
\cite{kogan}
with whose help
one can show that $\Delta f_{\alpha}(t) = \Delta f_{\alpha}$ for $t > 0$.
Thus the intrinsic time dependence of $\Delta f^{(2)}(t)$ is
not revealed through this quantity.
[Note: a remark in Green, Ref.
\onlinecite{fg1},
that $\Delta f(t)$ is inherently time-dependent, is true
only for collision-time approximations.]
Eq. (\ref{rgf1}) implies that
the total fluctuation strength is constant:
${\sum}_{\bf r} \Omega({\bf r})
\langle \Delta f ({\bf r}, t) \rangle = \Delta N$.

\subsection{Dynamic Correlations}

We move to the frequency domain.
An important outcome of this analysis is the extension of
the fluctuation-dissipation (FD) relation to the nonequilibrium regime.
This requires expressing both the difference function $g$
and the adiabatic Boltzmann-Green function $G$
in terms of the dynamical response.
The Fourier transform ${\rm R}(\omega)$ of the time-dependent BGF
satisfies

\begin{equation}
\sum_{\beta}
{\left\{ B[W^A f]_{\alpha \beta}
- i\omega \delta_{\alpha \beta} \right\}}
{\rm R}_{\beta \alpha'}(\omega) =
\delta_{\alpha \alpha'},
\label{AXX}
\end{equation}

\noindent
making ${\rm R}(\omega)$ the resolvent for the
linearised operator of Eq. (\ref{bop}).
The global condition on ${\rm R}(\omega)$ from Eq. (\ref{rgf1}) is

\begin{equation}
\sum_{\alpha} {\rm R}_{\alpha \alpha'}(\omega)
= -{1\over {i(\omega + i\eta)} }, {~~~~~~ }\eta \to 0^+.
\label{rgf5}
\end{equation}

\noindent
At first sight this fails to match the corresponding
condition on the adiabatic Boltzmann-Green function,
Eq. (\ref{dgauss}).
To determine the solution of Eq. (\ref{AXG})
for $G$ in terms of the resolvent, we follow Kogan and Shul'man
\cite{kogan}
and introduce the intrinsically correlated part of
${\rm R}(\omega)$, namely

\begin{equation}
{\rm C}_{\alpha \alpha'}(\omega) = {\rm R}_{\alpha \alpha'}(\omega)
+ {1\over {i(\omega + i\eta)} }
{ {{\Delta f}_{\alpha}}\over {\Delta N} }.
\label{f2.2}
\end{equation}

\noindent
This correlated propagator satisfies a pair of
identities in the frequency domain;
\cite{kogan}
the transform of the relation $\Delta f(t) = \theta(t)\Delta f$ leads to

\begin{mathletters}
\label{cckt}
\begin{equation}
\sum_{\alpha'} {\rm C}_{\alpha \alpha'}(\omega)
\Delta f_{\alpha'} = 0
\label{cckta}
\end{equation}

\noindent
while Eq. (\ref{rgf5}) leads to

\begin{equation}
\sum_{\alpha} {\rm C}_{\alpha \alpha'}(\omega) = 0.
\label{ccktb}
\end{equation}
\end{mathletters}

\noindent
The second of these corresponds to the constraint on $G$.
Like ${\rm R}(\omega)$, the correlated BGF
is analytic in the upper half-plane ${\rm Im}\{\omega\} > 0$,
and satisfies the Kramers-Kr\"onig dispersion relations.
Unlike ${\rm R}(\omega)$, however, ${\rm C}(\omega)$
is regular for $\omega \to 0$.

We now obtain $g$ and $G$ in terms of the
correlated dynamical propagator. Consider the equation

\begin{equation}
\sum_{\alpha'}
  {\left\{
B[W^A f]_{\alpha \alpha'} - i\omega \delta_{\alpha \alpha'}
  \right\}}
{\rm g}_{\alpha'}(\omega)
= { { e{\bf {\widetilde E}}({\bf r}) }\over {\hbar} }{\bbox \cdot}
  { {\partial f^{\rm eq}_{\alpha}}\over {\partial {\bf k}} }
+ \sum_{\alpha'}
{g}_{\alpha} W^A_{\alpha \alpha'}
{g}_{\alpha'},
\label{cvlv1}
\end{equation}

\noindent
with solution

\begin{equation}
{\rm g}_{\alpha}(\omega)
= \sum_{\alpha'}
{\rm C}_{\alpha \alpha'}(\omega)
{ e{\bf {\widetilde E}} ({\bf r'})\over \hbar } {\bbox \cdot}
{ {\partial f^{\rm eq}_{\alpha'}}\over {\partial {\bf k'}} }
+ \sum_{\alpha' \beta}
{\rm C}_{\alpha \alpha'}(\omega)
{g}_{\alpha'} W^A_{\alpha' \beta} {g}_{\beta}.
\label{AX15.3}
\end{equation}

\noindent
The uncorrelated component of ${\rm R}(\omega)$
does not contribute to the right-hand side of this equation;
in the first term it results in a decoupling of
the summation over ${\alpha'}$, yielding zero
because $\partial f^{\rm eq}_{\alpha'}/\partial {\bf k'}$
is odd in ${\bf k'}$.
In the second term, decoupling means that
the double summation over ${\alpha'}$ and ${\beta}$
vanishes by antisymmetry.
In the static limit Eq. (\ref{cvlv1}) becomes
the inhomogeneous equation (\ref{AX5}),
and moreover ${\rm g}(0)$ satisfies Eq. (\ref{gauss}),
the sum rule for $g$. Therefore

\begin{equation}
g_{\alpha}
= \sum_{\alpha'}
{\rm C}_{\alpha \alpha'}(0)
{ e{\bf {\widetilde E}} ({\bf r'})\over \hbar } {\bbox \cdot}
{ {\partial f^{\rm eq}_{\alpha'}}\over {\partial {\bf k'}} }
+ \sum_{\alpha' \beta}
{\rm C}_{\alpha \alpha'}(0) g_{\alpha'} W^A_{\alpha' \beta} g_{\beta}.
\label{cvlv2}
\end{equation}

\noindent
This identity is central to the FD relation.

In models with symmetric scattering
$W^A$ is zero and the adiabatic BGF assumes
a simple form on varying both sides of Eq. (\ref{cvlv2}):

\begin{equation}
G_{\alpha \alpha'} = {\rm C}_{\alpha \alpha'}(0)
{ e{\bf {\widetilde E}} ({\bf r'})\over \hbar } {\bbox \cdot}
{ {\partial}\over {\partial {\bf k'}} }.
\label{AXG2}
\end{equation}

\noindent
More generally, an analysis similar to that for
${\rm g}(\omega)$ can be used directly
for the adiabatic propagator.
Introduce the operator ${\rm G}(\omega)$, defined to
satisfy the dynamic extension of Eq. (\ref{AXG}),

\begin{equation}
\sum_{\beta}
  {\left\{
B[W^A f]_{\alpha \beta} - i\omega \delta_{\alpha \beta} 
  \right\}}
{\rm G}_{\beta \alpha'}(\omega)
=
\delta_{\alpha \alpha'}
{\left[
{ { e{\bf {\widetilde E}} ({\bf r'})}
\over {\hbar} }{\bbox \cdot}
   { {\partial }\over {\partial {\bf k'}} }
+ \sum_{\beta} W^A_{\beta \alpha'}
{g}_{\beta}
\right]} - W^A_{\alpha \alpha'}
{g}_{\alpha}.
\label{AYG}
\end{equation}

\noindent
This has the solution

\begin{equation}
{\rm G}_{\alpha \alpha'}(\omega)
= {\rm C}_{\alpha \alpha'}(\omega)
{ { e{\bf {\widetilde E}} ({\bf r'})}
\over {\hbar} }{\bbox \cdot}
   { {\partial }\over {\partial {\bf k'}} }
- \sum_{\beta}
{\Big(
{\rm C}_{\alpha \alpha'}(\omega)
-
{\rm C}_{\alpha \beta}(\omega)
\Big)}
W^A_{\alpha' \beta} {g}_{\beta}.
\label{cvlv4}
\end{equation}

\noindent
In the first term on the right-hand side, the uncorrelated
component of ${\rm R}(\omega)$ makes no contribution  after decoupling
because the physical distributions $F_{\alpha}$
on which ${\rm G}(\omega)$ operates vanish sufficiently fast that
${\sum}_{\bf k} {\partial F}_{\alpha}/{\partial {\bf k}} = {\bf 0}$.
In the second right-hand term the uncorrelated parts of
${\rm R}_{\alpha \alpha'}$ and ${\rm R}_{\alpha \beta}$
cancel directly. We conclude as before that

\begin{equation}
G_{\alpha \alpha'}
= {\rm C}_{\alpha \alpha'}(0)
{ { e{\bf {\widetilde E}} ({\bf r'})}
\over {\hbar} }{\bbox \cdot}
   { {\partial }\over {\partial {\bf k'}} }
- \sum_{\beta}
{\Big(
{\rm C}_{\alpha \alpha'}(0)
-
{\rm C}_{\alpha \beta}(0)
\Big)}
W^A_{\alpha' \beta} g_{\beta}.
\label{cvlv5}
\end{equation}

\noindent
This is a crucial result because it shows
(a) that the adiabatic structure of the steady state, through $G$,
is of one piece with the dynamics, and
(b) that the nonequilibrium correlations originate {\em manifestly}
from the equilibrium state, through $G \Delta f^{\rm eq}$.
We have thus proved that the kinetic BGF analysis is formally
self-sufficient once its boundary conditions are given.
This fact is embodied in the frequency sum rule

\begin{equation}
{{\delta g_{\alpha}}\over {\delta f^{\rm eq}_{\alpha'}}}
{~}={~} {2\over \pi}\int^{\infty}_0 {d\omega\over \omega}
{\rm Im} {\{ {\rm G}_{\alpha \alpha'}(\omega) \}}.
\label{fsrg}
\end{equation}

The vehicle for the physics of current noise is
the velocity autocorrelation.
It is a two-point distribution in real space, built on
the correlated part of the two-particle fluctuation
$\Delta {\rm f}^{(2)}_{\alpha \alpha'}
= {\rm R}_{\alpha \alpha'}\Delta f_{\alpha'}$
and taking the form
\cite{ggk}

\begin{equation}
{ \langle\!\langle {\bf v} {\bf v'}
{\Delta {\rm f}^{(2)} } ({\bf r}, {\bf r'}; \omega) 
\rangle\!\rangle}_c'
\buildrel \rm def \over =
{1\over \Omega({\bf r})} {1\over \Omega({\bf r'})}
\sum_{{\bf k}, s} \sum_{{\bf k'}, s'}
{\bf v}_{{\bf k} s}
 {{\rm Re} \{{\rm C}_{\alpha \alpha'}(\omega)\} } 
{\bf v}_{{\bf k'} s'}
\Delta f_{\alpha'}.
\label{AXY}
\end{equation}

\noindent
The nonlocal velocity autocorrelation
provides the direct basis for shot-noise calculations
when the distance $|{\bf r} - {\bf r'}|$ becomes
comparable to the mean free path.
The local function derived from it,

\begin{equation}
S_f ({\bf r}, \omega)
= e^2 \sum_{\bf r'} \Omega({\bf r'})
{\langle\!\langle 
( {\bf {\widetilde E}}({\bf r}){\bbox \cdot}{\bf v} )
( {\bf {\widetilde E}}({\bf r'}){\bbox \cdot}{\bf v'} )
{\Delta {\rm f}^{(2)}({\bf r}, {\bf r'}; \omega) }
\rangle\!\rangle}_c',
\label{Svv}
\end{equation}

\noindent
has a macroscopic reach since in effect it samples fluctuations
over the bulk.
It is closely related to the current-noise spectral density
and satisfies the nonequilibrium FD relation
discussed in Sec. III.

\subsection{Coulomb Effects}

We generate the Boltzmann-Green functions
in the presence of induced fluctuations of the electric fields.
Variations are now unrestricted.
The resulting description of screening effects extends the
physics of classical space-charge suppression.
\cite{suppr}
We consider samples with a fixed dielectric constant, and
likewise for the reservoirs.

The equation for the screened equilibrium fluctuation
${\widetilde \Delta} f^{\rm eq} \equiv k_BT \delta f^{\rm eq}/\delta \mu$
is obtained by operating on Eq. (\ref{AX1.1})
satisfied by the equilibrium distribution. We have

\begin{equation}
  { {\bf v}_{{\bf k} s } }{\bbox \cdot} 
{ { \partial {\widetilde {\Delta}} f^{\rm eq}_{\alpha} }
\over { \partial {\bf r} } }
- { { e{\bf E}_0({\bf r}) }\over {\hbar} } {\bbox \cdot}
{ { \partial {\widetilde {\Delta}} f^{\rm eq}_{\alpha} }
\over { \partial {\bf k} } }
=
{\left(
{e \over \hbar} \sum_{\alpha'}
{ {\delta {\bf E}_0({\bf r})}\over {\delta f^{\rm eq}_{\alpha'}} }
{\widetilde {\Delta}} f^{\rm eq}_{\alpha'}
\right)}
{\bbox \cdot} 
{{\partial f^{\rm eq}_{\alpha} }\over {\partial {\bf k}} }.
\label{scr5}
\end{equation}

\noindent
Detailed balance keeps the equation collisionless
while the Poisson equation (\ref{poissoneq}) implies that,
within the system boundaries,
the variation of $e{\bf E}_0$ with respect to $f^{\rm eq}$
is the Coulomb force for an electron,

\begin{equation}
e{ {\delta {\bf E}_0 ({\bf r})}
\over {\delta f^{\rm eq}_{\alpha'}} }
= -e{\bf E}_{C}({\bf r} - {\bf r'})
\equiv {{\partial}\over {\partial {\bf r}}}
V_{C}({\bf r} - {\bf r'})
\label{scr0}
\end{equation}

\noindent
where $V_{C}({\bf r}) = e^2/\epsilon |{\bf r}|$
is the Coulomb potential. As a result Eq. (\ref{scr5}) becomes

\begin{equation}
  { {\bf v}_{{\bf k} s } }{\bbox \cdot} 
{ { \partial {\widetilde {\Delta}} f^{\rm eq}_{\alpha} }
\over { \partial {\bf r} } }
- { { e{\bf E}_0({\bf r}) }\over {\hbar} } {\bbox \cdot}
{ { \partial {\widetilde {\Delta}} f^{\rm eq}_{\alpha} }
\over { \partial {\bf k} } }
= -{e \over \hbar} \sum_{\alpha'}
{{\partial f^{\rm eq}_{\alpha} }\over {\partial {\bf k}} }
{\bbox \cdot} 
{\bf E}_{C}({\bf r} - {\bf r'})
{\widetilde {\Delta}} f^{\rm eq}_{\alpha'}.
\eqnum{$\ref{scr5}'$}
\end{equation}

\noindent
Viewed as a variant of the equilibrium BTE,
Eq. ({$\ref{scr5}'$}) is inhomogeneous. Its solution includes a term
proportional to the homogeneous solution, which in this case
is the bare fluctuation $\Delta f^{\rm eq}$.
Let $\gamma_{C}$ be the proportionality constant. Then

\begin{equation}
{\widetilde {\Delta}} f^{\rm eq}_{\alpha}
= \gamma_{C} \Delta f^{\rm eq}_{\alpha}
- {e\over \hbar}\sum_{\alpha' \alpha''}
{\rm C}^{\rm eq}_{\alpha \alpha'}(0)
{ {\partial f^{\rm eq}_{\alpha'}}
\over {\partial {\bf k'}} } {\bbox \cdot}
{\bf E}_{C}({\bf r'} - {\bf r''})
{\widetilde {\Delta}} f^{\rm eq}_{\alpha''},
\label{scr1.1}
\end{equation}

\noindent
in which ${\rm C}^{\rm eq}$ is the correlated part of the resolvent
for the equilibrium state. The integral on the right-hand side of
Eq. (\ref{scr1.1}) has a structure similar to Eq. (\ref{cvlv2}), in that
the uncorrelated part of the resolvent
gives no contribution after
decoupling of the intermediate wave-vector sums.

\subsubsection{Thomas-Fermi Screening}

The constant $\gamma_{C}$ is sensitive to the physics of charge transfer
between sample and reservoirs.
Recall that the fluctuation $\Delta f^{\rm eq}_{\alpha} = k_BT
{\partial f^{\rm eq}_{\alpha}}/{\partial \phi_{\alpha}}$
is a measure of the electrons' response, as a Fermi liquid,
to a change in the effective Fermi level
$\phi_{\alpha} = \mu - V_0({\bf r})$. The latter is the net contribution
from kinematics {\em alone} to the cost
of adding an electron locally to the system; the electrostatic
energy $V_0({\bf r})$ is excluded from the Fermi-liquid accounting.

When the Coulomb fields are frozen, as in the restricted analysis,
the Fermi-level variation is that of the
global chemical potential: $\delta \phi_{\alpha} = \delta \mu$.
In the full Coulomb problem, we must offset the energy
cost of charge transfer from reservoir to sample.
The Coulomb energy needed to add an electron to the conductor is

\begin{equation}
u_{c} = {1\over N} \sum_{\alpha} V_0({\bf r}) f^{\rm eq}_{\alpha}.
\label{scru}
\end{equation}

\noindent
A corresponding term $u_r$ characterises the reservoirs. However, $u_r$
cannot be probed directly; its effects are absorbed within the operational
definition of the chemical potential. This means that $u_r = 0$ identically,
and that $u_{c}$ thus represents the net work
to move an electron from reservoir to sample.
It is the conduction-electron contribution
to the contact potential.\cite{ashmer}
(By contrast, the core-electron contribution determines
the offset in the band bottom
$\varepsilon_s({\bf k} \!\! = \!\! {\bf 0}; {\bf r})$. This is
independent of $\mu$ and does not appear explicitly in the
variational analysis.)

It follows that the portion of the chemical potential
sustaining the Fermi liquid in the conductor is $\mu - u_{c}$.
Free variation of the global parameter $\mu$
generates the coefficient

\begin{equation}
\gamma_{C}
=  {\delta\over {\delta \mu}}(\mu - u_{c})
= 1 - {{\delta u_{c}}\over {\delta \mu}}.
\label{scru1}
\end{equation}

\noindent
Application to Eq. (\ref{scr1.1}) 
of Eq. (\ref{ccktb}), that is
the sum rule ${\sum}_{\alpha} {\rm C}_{\alpha \alpha'} = 0$,
establishes the normalisation

\begin{equation}
\sum_{\alpha} {\widetilde {\Delta}} f^{\rm eq}_{\alpha}
= \gamma_{C} \Delta N.
\label{scrc}
\end{equation}

\noindent
If sample and reservoir have matching
electronic properties, then $u_{c} = u_r \equiv 0$ and $\gamma_{C} = 1$.
This is the norm for noise measurements in metallic wires.
If, on the other hand, the sample differs substantially
from the reservoir in metallic structure,
then from Eq. (\ref{scru}) and the form for
$V_0({\bf r})$ as the solution to the Poisson equation
(\ref{poissoneq}) one obtains

\begin{equation}
\gamma_{C}
= 1 - {1\over N} \sum_{\alpha}
{\left(
\sum_{\alpha'} V_{C}({\bf r} - {\bf r'}) f^{\rm eq}_{\alpha'}
+ V_0({\bf r}) - {u_{c}\over N}
\right)}
{ {{\widetilde {\Delta}} f^{\rm eq}_{\alpha}}\over k_BT },
\label{scruc}
\end{equation}

\noindent
an instance of suppression by self-consistent Thomas-Fermi screening.

There is strong indirect evidence that this mechanism is the
major determinant of low-noise performance in heterojunction
field-effect devices.
\cite{gc}
At mesoscopic scales, one can anticipate a wide variety of
interfacial screening behaviours for the fluctuations.
Potentially interesting is the case of metal--semiconductor--metal
structures. See Appendix \ref{apxcsd}, Eq. (\ref{apxcsd11}).

Note that $\gamma_{C}$ enters only at the two-body level. It cannot
renormalise the one-body distribution $g$, or any averages constructed
with $g$. This includes transport coefficients such as the mobility.

Equation (\ref{scr1.1}) is solved by introducing
a Coulomb screening operator $\Gamma^{\rm eq}(0)$, whose
inverse is

\begin{mathletters}
\label{scr1.2}
\begin{equation}
{\Bigl( \Gamma^{\rm eq}(0)^{-1} \Bigr)}_{\alpha \alpha'}
= \delta_{\alpha \alpha'}
+ {e\over \hbar}
\sum_{\beta}
{\rm C}^{\rm eq}_{\alpha \beta}(0)
{ {\partial f^{\rm eq}_{\beta}}
\over {\partial {\bf k_{\beta}}} } {\bbox \cdot}
{\bf E}_{C}({\bf r}_{\beta} - {\bf r'}).
\label{scr1.2a}
\end{equation}

\noindent
This yields

\begin{equation}
{\widetilde {\Delta}} f^{\rm eq}_{\alpha}
= \sum_{\alpha'}
\Gamma^{\rm eq}_{\alpha \alpha'}(0)
{\Bigl( \gamma_{C} \Delta f^{\rm eq}_{\alpha'} \Bigr)},
\label{scr1.2b}
\end{equation}
\end{mathletters}

\noindent
analogous to the Lindhard screening theory
of the electron gas in the static limit.
\cite{nozpin}
Taken together with Eq. (\ref{scruc}),
it allows a closed-form solution for $\gamma_{C}$.

\subsubsection{Collision-Mediated Screening}

Away from equilibrium we define the Coulomb
screening operator through its inverse

\begin{equation}
{\Bigl( \Gamma(\omega)^{-1} \Bigr)}_{\alpha \alpha'}
\buildrel \rm def \over =
\delta_{\alpha \alpha'}
+ {e\over \hbar} \sum_{\beta}
{\rm C}_{\alpha \beta}(\omega)
{ {\partial f_{\beta}}
\over {\partial {\bf k_{\beta}}} } {\bbox \cdot}
{\bf E}_{C}({\bf r}_{\beta} - {\bf r'}).
\label{scr10.1}
\end{equation}

\noindent
While $\gamma_{C}$ is collisionless,
the operator $\Gamma$ captures the dynamics of interaction
between scattering and screening, an exclusively
nonequilibrium process. A few elementary results
for both collisional and Thomas-Fermi screening
are discussed in Appendix \ref{apxcsd}.
An important property of the collision-mediated screening operator,
following from Eq. (\ref{ccktb}), is

\begin{equation}
\sum_{\alpha} \Gamma_{\alpha \alpha'}(\omega) = 1.
\label{gprop}
\end{equation}

The significance of $\Gamma$ first becomes evident
in obtaining the screened adiabatic propagator
${\widetilde G}_{\alpha \alpha'} =
\delta g_{\alpha}/\delta f^{\rm eq}_{\alpha'}$,
whose unrestricted BTE [cf Eq. (\ref{AXG}) for $G$] is

\begin{eqnarray}
\sum_{\beta}
B[W^A f]_{\alpha \beta} {\widetilde G}_{\beta \alpha'}
=&& \sum_{\beta}
B[W^A f]_{\alpha \beta} G_{\beta \alpha'}
+  {e\over \hbar}
{\left(
 \sum_{\beta}
 { {\delta {\bf E} ({\bf r}) }\over {\delta f_{\beta}} } 
 { {\delta f_{\beta}}\over
 { {\delta f^{\rm eq}_{\alpha'} }} }
\right)}
{\bbox \cdot} { {\partial f_{\alpha} } \over {\partial {\bf k}} }
\cr
&&- {e\over \hbar}
{ {\delta {\bf E}_0 ({\bf r}) } \over
{ {\delta f^{\rm eq}_{\alpha'} }} }
{\bbox \cdot} { {\partial f^{\rm eq}_{\alpha} }
		       \over {\partial {\bf k}} }
\label{scr1}
\end{eqnarray}

\noindent
Poisson's equation (\ref{poiss}) once again determines
the variation of ${\bf E}$ with respect to $f$ as

\begin{equation}
{ {\delta {\bf E}({\bf r})}\over {\delta f_{\alpha'}} }
= -{\bf E}_{C}({\bf r} - {\bf r'}).
\label{scr2}
\end{equation}

\noindent
The solution of Eq. (\ref{scr1}) is a two-pass process,
in which one first resolves the Boltzmann operator $B[W^A f]$
and then invokes $\Gamma$ to rationalise the transformed equation:

\begin{eqnarray}
{\widetilde G}_{\alpha \alpha'}
=&& 
G_{\alpha \alpha'} - {e\over \hbar}\sum_{\beta \beta'}
{\rm C}_{\alpha \beta}(0)
{ {\partial f_{\beta}}\over {\partial {\bf k}_{\beta}} } {\bbox \cdot}
{\bf E}_{C}({\bf r}_{\beta} - {\bf r}_{\beta'})
\Bigl( \delta_{\beta' \alpha'} +
{\widetilde G}_{\beta ' \alpha'} \Bigr)
\cr
&&+ {e\over \hbar}\sum_{\beta}
{\rm C}_{\alpha \beta}(0)
{ {\partial f^{\rm eq}_{\beta}}\over
  {\partial {\bf k}_{\beta}} } {\bbox \cdot}
{\bf E}_{C}({\bf r}_{\beta} - {\bf r'})
\cr
{\left. \right.} \cr
=&&
\sum_{\beta}
\Gamma_{\alpha \beta}(0)
{\left( G_{\beta \alpha'}
- {e\over \hbar} \sum_{\beta'}
{\rm C}_{\beta \beta'}(0)
{ {\partial g_{\beta'}}\over {\partial {\bf k}_{\beta'}} }
{\bbox \cdot} {\bf E}_{C}({\bf r}_{\beta'} - {\bf r'})
\right)}.
\label{scr3}
\end{eqnarray}

\noindent
We have also used
${\delta f_{\beta'}}/{\delta f^{\rm eq}_{\alpha'}}
= \delta_{\beta' \alpha'} + {\widetilde G}_{\beta' \alpha'}$.
The screened steady-state fluctuation is now

\begin{equation}
{\widetilde {\Delta}} f_{\alpha}
= {\widetilde {\Delta}} f^{\rm eq}_{\alpha}
+ \sum_{\alpha'} {\widetilde G}_{\alpha \alpha'}
{\widetilde {\Delta}} f^{\rm eq}_{\alpha'}.
\label{scrg.1}
\end{equation}

\noindent
The main outcome of the structure of ${\widetilde G}$
is the invariance of the fluctuation strength over the sample;
this follows from
${\sum}_{\alpha} {\widetilde G}_{\alpha \alpha'} = 0$.
Thus

\begin{equation}
\sum_{\alpha} {\widetilde {\Delta}} f_{\alpha}
= \gamma_{C} \Delta N
\equiv  {\widetilde {\Delta}} N.
\label{scrg.2}
\end{equation}

\subsubsection{Dynamics}

We now examine the screened dynamics.
In the time domain the screened resolvent
${\widetilde R}(t)$ has the same
formal definition, Eq. (\ref{AX8}),  as its restricted analogue.
The unrestricted equation of motion, Fourier transformed, is

\begin{equation}
\sum_{\beta}
{\left\{ B[W^A f]_{\alpha \beta}
- i\omega \delta_{\alpha \beta} \right\}}
{\rm {\widetilde R}}_{\beta \alpha'}(\omega) =
\delta_{\alpha \alpha'}
- {e\over \hbar}
\sum_{\beta}
{ {\partial f_{\alpha} } \over {\partial {\bf k}} } {\bbox \cdot}
{\bf E}_{C}({\bf r} - {\bf r}_{\beta})
{\rm {\widetilde R}}_{\beta \alpha'}(\omega),
\label{scr7}
\end{equation}

\noindent
with solution

\begin{equation}
{\rm {\widetilde R}}_{\alpha \alpha'}(\omega)
= \sum_{\beta} \Gamma_{\alpha \beta}(\omega)
{\rm R}_{\beta \alpha'}(\omega).
\label{scr8}
\end{equation}

\noindent
This resolvent obeys identities analogous to those for
${\rm R}(\omega)$, namely Eq. (\ref{rgf5}) and, in the time domain,
${\widetilde R}_{\alpha \alpha'}(t \to \infty)
= {\widetilde {\Delta}} f_{\alpha}/{\widetilde {\Delta}} N.$
Together with Eq. (\ref{scrg.2}),
equality of the residues at $\omega = 0$ on each side of Eq. (\ref{scr8})
implies the relation

\begin{equation}
{\widetilde {\Delta}} f_{\alpha}
= \gamma_{C}
\sum_{\alpha'} \Gamma_{\alpha \alpha'}(0) \Delta f_{\alpha'},
\label{scr10}
\end{equation}

\noindent
equivalent to Eq. (\ref{scrg.1}) by the properties of $G$ and $B[W^Af]$.

The correlated propagator, with screening, is

\begin{equation}
{\rm {\widetilde C}}_{\alpha \alpha'}(\omega)
= {\rm {\widetilde R}}_{\alpha \alpha'}(\omega)
+ { 1\over {i(\omega + i\eta)} }
{ {{\widetilde {\Delta}} f_{\alpha}}\over {{\widetilde {\Delta}} N} }.
\label{scr4.3}
\end{equation}

\noindent
Its structure follows from combining Eqs.
(\ref{f2.2}), (\ref{scr8}), and (\ref{scr10}) for

\begin{equation}
{\rm {\widetilde C}}_{\alpha \alpha'}(\omega)
= \sum_{\beta} \Gamma_{\alpha \beta}(\omega) 
{\rm C}_{\beta \alpha'}(\omega)
- \gamma_{C} \sum_{\beta}
{\left(
{ { \Gamma_{\alpha \beta}(\omega)
  - \Gamma_{\alpha \beta}(0) } \over { i\omega} }
\right)}
{ {\Delta f_{\beta}} \over {{\widetilde {\Delta}} N} }.
\label{scr11}
\end{equation}

\noindent
Applying the screened form of Eq. (\ref{cckta}), that is
${\sum}_{\beta'} {\rm {\widetilde C}}_{\beta \beta'}(\omega)
{\widetilde {\Delta}} f_{\beta'} = 0$,
to Eq. (\ref{scr11}) generates

\[
\sum_{\beta \beta'} \Gamma_{\alpha \beta}(\omega)
{\rm C}_{\beta \beta'}(\omega)
{\widetilde {\Delta}} f_{\beta'}
= \gamma_{C} \sum_{\beta}
{\left(
{ { \Gamma_{\alpha \beta}(\omega)
  - \Gamma_{\alpha \beta}(0) } \over { i\omega} }
\right)} \Delta f_{\beta}.
\]

\noindent
Fed back into Eq. (\ref{scr11}), this produces

\begin{equation}
{\rm {\widetilde C}}_{\alpha \alpha'}(\omega)
= 
\sum_{\beta \beta'}
\Gamma_{\alpha \beta}(\omega)
{\rm C}_{\beta \beta'}(\omega)
{\left(
\delta_{\beta' \alpha'} - { { {\widetilde {\Delta}} f_{\beta'} }
                               \over {{\widetilde {\Delta}} N} }
\right)}.
\label{scr15}
\end{equation}

\subsubsection{Current-Current Correlation}

All the components are in place to construct the velocity
autocorrelation function in the presence of screening.
At zero frequency this is

\begin{equation}
{ \langle\!\langle {\bf v} {\bf v'}
{ {\widetilde \Delta} {\rm f}^{(2)} }({\bf r}, {\bf r'}; 0) 
\rangle\!\rangle}_c'
\buildrel \rm def \over =
{1\over \Omega({\bf r})} {1\over \Omega({\bf r'})}
\sum_{{\bf k}, s} \sum_{{\bf k'}, s'}
{\bf v}_{{\bf k} s}
{\rm {\widetilde C}}_{\alpha \alpha'}(0)
{\bf v}_{{\bf k'} s'}
{\widetilde \Delta} f_{\alpha'}.
\label{scr16}
\end{equation}

\noindent
At finite frequency we must add the displacement-current
contribution to the fluctuations. The velocity is replaced
with the nonlocal operator

\begin{equation}
{\bf u}_{{\bf k} s}({\bf r}, {\bf r''}; \omega)
\equiv 
{{\delta_{{\bf r} {\bf r''}}}\over \Omega({\bf r})}
{\bf v}_{{\bf k} s}
- {{i\omega \epsilon}\over {4\pi e}} {\bf E}_{C}({\bf r} - {\bf r''}),
\label{displ}
\end{equation}

\noindent
which requires two intermediate spatial sums to be
incorporated within the expectation
$\langle\!\langle
{\rm Re} {\{\bf u} {\widetilde \Delta} {\rm f}^{(2)} {\bf u'}^*\}
\rangle\!\rangle'_c$. For $\omega = 0$ this recovers Eq. (\ref{scr16}),
a more complex expression than its bare counterpart
Eq. (\ref{AXY}). In practice, collisional Coulomb effects are
dominant in mesoscopic and in strongly inhomogeneous systems.
\cite{reggi2,rggcoul}

\section{Applications}

The first of our applications connects thermal fluctuations
and dissipation in the bulk nonequilibrium context.
Little is known of the effects of degeneracy
on noise beyond the linear limit,
\cite{mc}
and we analyse them here.
In our second application we investigate the many-body nature of
mesoscopic shot noise. For degenerate electrons we show that
thermal and shot noise have very different physical properties
not easily subsumed under a single formula.
\cite{theor}

\subsection{Nonequilibrium Fluctuation-Dissipation Relation}

The fluctuation-dissipation relation near equilibrium connects
the spectral density of the thermal fluctuations to the
dissipative effects of the steady current in the system.
However, dissipation by itself does not exhaust the physics of
this sum rule. There are nonlinear terms,
negligible in linear response, that
dominate the high-field behaviour of the current noise.
\cite{sw2,gc}
We calculate these contributions.
Since the relation is macroscopic, to lowest order
we omit Coulomb screening effects; these are
weak in the bulk metallic limit
(see for example Appendix \ref{apxcsd}). 

The resolvent property of ${\rm R}(\omega)$
provides a formal connection between the
steady-state (one-body) solution $g$ and
the dynamical (two-body) fluctuation
$\Delta {\rm f}^{(2)}$ at the semiclassical level.
Taken to its equilibrium limit this becomes the familiar theorem.
\cite{nozpin}
The connection is made in two steps.
Consider the kinematic identity

\begin{equation}
{ {\partial f^{\rm eq}_{\alpha}}\over {\partial {\bf k}} }
= -{ \hbar \over {k_BT} }
{\bf v}_{{\bf k} s}
\Delta f^{\rm eq}_{\alpha}
\label{AX14}
\end{equation}

\noindent
and apply it to the leading term
on the right-hand side of Eq. (\ref{cvlv2}). The result is

\begin{equation}
g_{\alpha} = -{ {e}\over {k_B T} }
\sum_{\alpha'}
{\rm C}_{\alpha \alpha'}(0)
( {\bf {\widetilde E}} {\bbox \cdot} {\bf v} )_{\alpha'}
\Delta f^{\rm eq}_{\alpha'} + h_{\alpha},
\label{AX15}
\end{equation}

\noindent
in which $h_{\alpha} = {\sum}_{\alpha' \beta}
{\rm C}_{\alpha \alpha'}(0) g_{\alpha'} W^A_{\alpha' \beta} g_{\beta}$.
Evaluating the current density according to
${\bf J}({\bf r}) = -e\langle {\bf v} g \rangle$, the power density
$P({\bf r}) = {\bf {\widetilde E}} ({\bf r}) {\bbox \cdot} 
{\bf J}({\bf r})$ for Joule heating can be written as

\begin{equation}
P({\bf r}) =
{ {e^2}\over {k_B T} }
{1\over \Omega({\bf r})}
\sum_{{\bf k},s}
\sum_{\alpha'}
( {\bf {\widetilde E}} {\bbox \cdot} {\bf v} )_{\alpha}
{\rm C}_{\alpha \alpha'}(0)
( {\bf {\widetilde E}} {\bbox \cdot} {\bf v} )_{\alpha'}
\Delta f^{\rm eq}_{\alpha'}
- e{\langle {\bf {\widetilde E}} {\bbox \cdot} {\bf v} h \rangle}.
\label{AY16}
\end{equation}

In the second step we take the one-point spectral
function $S_f$ in the static limit,
substituting for $\Delta f$ from Eq. (\ref{AX7})
in the right-hand side of Eq. (\ref{Svv}) to give

\begin{equation}
S_f ({\bf r}, 0) =
{e^2\over \Omega({\bf r})}
\sum_{{\bf k},s}
\sum_{\bf r'} \sum_{{\bf k'},s'}
( {\bf {\widetilde E}} {\bbox \cdot} {\bf v} )_{\alpha}
{\rm C}_{\alpha \alpha'}(0)
( {\bf {\widetilde E}} {\bbox \cdot} {\bf v} )_{\alpha'}
\Delta f^{\rm eq}_{\alpha'}
+ S_g ({\bf r}, 0),
\label{eq17}
\end{equation}

\noindent
where $S_g({\bf r}, 0)$ is generated by replacing
$\Delta f$ with $\Delta g = \sum G \Delta f^{\rm eq}$
in Eq. (\ref{AXY}), and subsequently in Eq. (\ref{Svv}).
Direct comparison of Eqs. (\ref{AY16}) and (\ref{eq17}) leads to

\begin{equation}
{ {S_f ({\bf r}, 0)}\over {k_B T} }
= P({\bf r})
+ e{\langle {\bf {\widetilde E}} {\bbox \cdot} {\bf v} h \rangle}
+ { {S_g ({\bf r}, 0)}\over {k_B T} }.
\label{eq18}
\end{equation}

\noindent
This is the nonequilibrium FD relation.

The standard linear-response result follows.
The term in $h$ on the right-hand side
varies as ${\widetilde E} g^2$,
while the final term varies as
${\widetilde E}^2 \Delta g$; therefore
both of these contributions are of order
${\widetilde E}^3$.
Suppose that the system is homogeneous and that
${\bf {\widetilde E}} = {\bf E}$ acts along the {\em x}-axis:
then division by $E^2$
on both sides of Eq. (\ref{eq18}) gives

\begin{equation}
{1\over E^2}
{ {S_f ({\bf r}, 0)}\over k_BT}
{~\rightarrow~} {|J_x|\over E}
= \sigma,
\label{eq19}
\end{equation}

\noindent
where $\sigma$ is the low-field conductivity.
Eq. (\ref{eq19}) is the near-equilibrium statement.

The purely nonequilibrium structures beyond $P({\bf r})$
can be expanded similarly to it. We discuss the symmetric-scattering case,
for which there is no contribution
$e\langle {\bf {\widetilde E}} {\bbox \cdot} {\bf v} h \rangle$.
Within $S_g$ we apply the formula for the adiabatic 
Boltzmann-Green function, Eq. (\ref{AXG2}),
to express $\Delta g$
in terms of the correlated propagator
${\rm C}$. This produces two equivalent closed forms
for the higher-order correlation:

\begin{mathletters}
\label{eq22}
\begin{equation}
S_g ({\bf r}, 0)
= {e^2\over \Omega({\bf r})} \sum_{{\bf k}, s} 
\sum_{\beta}
( {\bf {\widetilde E}} {\bbox \cdot} {\bf v} )_{\alpha}
{\rm C}_{\alpha \beta}(0)
( {\bf {\widetilde E}} {\bbox \cdot} {\bf v} )_{\beta}
\sum_{\alpha'}
{\rm C}_{\beta \alpha'}(0)
{ {e {\bf {\widetilde E}}} ({\bf r'}) \over \hbar} {\bbox \cdot}
{ {\partial \Delta f^{\rm eq}_{\alpha'}}\over
{\partial {\bf k'}} },
\label{eq22a}
\end{equation}
\begin{equation}
S_g ({\bf r}, 0)
= -{e^3\over {k_BT}} {1\over \Omega({\bf r})}
\sum_{{\bf k}, s}
\sum_{\alpha'}
( {\bf {\widetilde E}} {\bbox \cdot} {\bf v} )_{\alpha}
   {
(
{\rm C}(0)
{\bf {\widetilde E}} {\bbox \cdot} {\bf v}
)
}^2_{\alpha \alpha'}
(1 - 2f^{\rm eq}_{\alpha'}) \Delta f^{\rm eq}_{\alpha'}.
\label{eq22b}
\end{equation}
\end{mathletters}

\noindent
Equation (\ref{eq22b}) follows from (\ref{eq22a}) after
using Eq. (\ref{AX14}) to express
${\partial \Delta f^{\rm eq}}/{\partial {\bf k}}$
in terms of $f^{\rm eq}$ and $\Delta f^{\rm eq}$,
and absorbing an inner sum into
$( {\rm C} {\bf {\widetilde E}} {\bbox \cdot} {\bf v} )^2$.

The term above is markedly different from
the rate of energy loss $P({\bf r})$ from Joule heating.
By contrast, $S_g ({\bf r}, 0)$ relates directly to
nonequilibrium broadening of the fluctuations, due to
the kinetic energy gained during ballistic motion;
the extent of the broadening is dynamically constrained by dissipation.
The impact of this term on current noise is felt only for
significant departures from equilibrium.

In a degenerate system there is an additional,
purely kinematic, constraint on field-driven broadening,
seen directly in the factor $(1 - 2f^{\rm eq})$
of Eq. (\ref{eq22b}). This inhibits the
contribution of $S_g$ relative to the
corresponding classical result, in which the
factor is unity. Suppression of electron heating
by Fermi-Dirac statistics reflects the large energy cost of displacing
electrons deep inside the Fermi sea.

To highlight the difference between dissipative and hot-electron
terms it is instructive to revisit a simple example,
\cite{gc,mbix}
the uniform electron gas in the
constant collision time (Drude) approximation,
subject to a field ${\bf E} = -E{\bf {\hat x}}$.
Expressions for the power density $P$
and hot-electron component $S_g$ are derived in Appendix {\ref{apxT}}.
The thermally driven current-current spectral density,
over a uniform sample of length $L_x$ and total volume $\Omega$, is
\cite{nougier}

\begin{eqnarray}
{\cal S}(E, \omega)
=&& 4 \sum_{\bf r}\Omega({\bf r}) \sum_{\bf r'}\Omega({\bf r'})
{\left\langle\!\!\left\langle
\left( -{ev_x\over L_x} \right) \left( -{ev'_x\over L_x} \right)
{\Delta {\rm f}^{(2)} } (\omega) 
\right\rangle\!\!\right\rangle}_c'
\cr
{\left. \right.} \cr
=&& 4 { {\Omega S_f(\omega)}\over {L_x^2 E^2} }.
\label{drude0}
\end{eqnarray}

\noindent
Introducing the conductance ${\cal G} = \Omega P/L_x^2 E^2$,
the static limit of the spectrum is determined by Eq. (\ref{eq18}):

\begin{equation}
{\cal S}(E,0) = 4{\cal G}k_BT
{\left[ 1 + { {S_g(0)}\over{Pk_BT} } \right]}
= 4{\cal G}k_BT
{\left[ 1 + {{\Delta {n}}\over {n}}
{\left( { {m^* \mu_e^2 E^2}\over k_BT } \right)} \right]}.
\label{drude1}
\end{equation}

\noindent
We have substituted for $P$ and $S_g$ from
Eqs. (\ref{apxT8}) and (\ref{apxT10}).
The electronic density is ${n}$ while
$\Delta {n} = \Delta N/\Omega$ is the number-fluctuation density.
The effective electron mass is $m^*$ and $\mu_e$ is the mobility.

The term $S_g/Pk_BT$ on the right-hand side
of Eq. (\ref{drude1}) is a relative measure of the hot-electron
contribution to the noise.
The inhibiting effect of degeneracy, through ${\Delta {n}}/{n}$,
is greatest at low temperature;
in terms of the Fermi energy $\varepsilon_F \propto {n}^{2/\nu}$
we have

\begin{equation}
{{\Delta {n}}\over {n}} = {k_BT\over {n}}
{{\partial {n}}\over {\partial \varepsilon_F}}
\to {{\nu k_BT}\over {2 \varepsilon_F}}.
\label{drude2}
\end{equation}

\noindent
When $\varepsilon_F \ll k_BT$ the ratio $\Delta n/n$ is unity;
the hot-electron term is that of a
classical electron gas (low density, high temperature), whose
high-field behaviour is ${\cal S} = 4{\cal G}m^*\mu_e^2E^2$
independently of $T$. On the other hand, when
$k_BT \ll \varepsilon_F$, the system is strongly degenerate
and Eq. (\ref{drude1}) with Eq. (\ref{drude2}) yields

\begin{equation}
{ {{\cal S}(E,0)}\over {4{\cal G}k_BT} }
\to 1 + {\nu\over 2}
{\left( { {m^* \mu_e^2 E^2}\over \varepsilon_F } \right)}.
\label{drude5}
\end{equation}

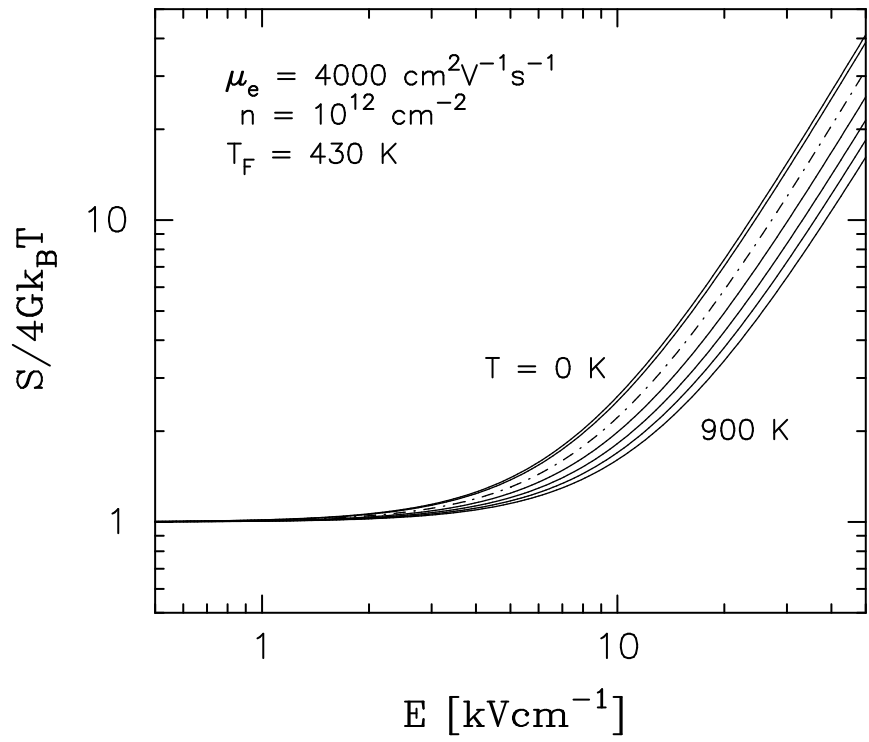
\begin{figure}
\caption{
Zero-frequency spectral density of nonequilibrium thermal noise
in a uniform, two-dimensional electron gas in GaAs, plotted
as a function of the external field from $T = 0$ to $900{\rm ~K}$
in $150{\rm ~K}$ steps.
Normalisation is to the equilibrium Johnson-Nyquist value.
At low temperature, degeneracy sets the scale of the contribution from
nonequilibrium electron heating.
At high temperature, the hot-electron component
shifts up towards higher fields as the equilibrium component
gains dominance. Dot-dashed line: thermal noise at $300{\rm ~K}$.
}
\label{fig1}
\end{figure}

\noindent
The thermal fluctuation spectrum necessarily
vanishes with temperature, but its ratio with the Johnson-Nyquist
spectral density $4{\cal G}k_B T$ continues to exhibit
a hot-electron excess, now scaled by the dominant energy
$\varepsilon_F$.
Figure \ref{fig1} illustrates the behaviour of the spectral ratio
for a two-dimensional electron gas, as a function of the applied field
as $T$ is taken from the degenerate limit to
above the Fermi temperature $T_F = \varepsilon_F/k_B$.
We see the gradual trend towards the
classical form of Eq. (\ref{drude1}) with rising temperature.

Equation (\ref{drude5}) may be compared with a
perturbative estimate by Landauer
\cite{landauer}
for the degenerate limit,
in which the analogous hot-electron contribution is
$(\delta U/ k_BT)^2$ where $\delta U \sim m^* \mu_e E v_F$
is a characteristic energy gain and $v_F$ is the Fermi velocity.
Taken at face value, this suggests that hot-electron effects
in the low-$T$ regime can be further enhanced by cooling.
A series expansion in powers of $E$ does not take into account
non-analyticity of the Boltzmann solutions
in the approach to equilibrium;
\cite{bakshi}
see also Eqs. (\ref{apxSN6.1}) and (\ref{apxSN13})
of our Appendix {\ref{apxSN}}.
Non-analyticity of the distribution function $f_{\bf k}$ 
precludes the reliable calculation of moment averages
by expanding away from equilibrium.

The relevance of non-analyticity to transport physics
has been questioned by Kubo, Toda, and Hashitsume.
\cite{kubo}
They ascribe its appearance to the simplistic treatment
of real collision processes by the Drude approximation,
despite strong evidence by Bakshi and Gross
\cite{bakshi}
that non-analyticity is generic to Boltzmann solutions.
Even in the Drude model, the nonperturbative solution
produces a physically coherent account of
the temperature dependence of nonequilibrium fluctuations,
while finite-order response theory does not.
Such clear qualitative differences
between perturbative and nonperturbative predictions
should be detectable in the nonequilibrium noise. 

There exist several alternative generalisations of the FD relation.
\cite{vvt,nougier,nerlul}
We mention the best known,
which defines the nonequilibrium noise temperature $T_n$ and
is pivotal to the interpretation of
device-noise data.
\cite{nougier}
This effective Nyquist temperature is obtained,
for a nonlinear operating point, by normalising ${\cal S}$ with
the  differential conductance ${\cal G}(E) = (e\Omega/L^2_x) \partial 
\langle v_x g \rangle / \partial E$ such that
$T_n(E) = {\cal S}(E,0)/4 {\cal G}(E) k_B$, corresponding
to the output of small-signal noise measurements.
Our Eqs. (\ref{AY16}) -- (\ref{eq22})
provide a microscopic framework for computing ${\cal S}$
in a wide class of degenerate systems. Since ${\cal G}(E)$ is also
calculable, this yields $T_n$.

\subsection{Shot Noise}

Carrier fluctuations manifest as shot noise when they
are induced by random changes in the discrete flux
at the terminals, rather than by thermal agitation distributed
through the body of the conductor. Consider an open segment of
electron gas between macroscopic leads.
For this segment we add up the transient,
time-of-flight correlations between the current at the source
boundary, $x = x_1$, and that at the drain boundary, $x = x_2$.

The total shot noise measured across the boundaries is the
resultant of two components. One component represents
the response at the drain terminal to the random {\em entry}
of electrons from the source reservoir, while the other represents
the response at the source terminal to the random {\em exit}
of electrons out to the drain reservoir. Thus

\begin{equation}
{\cal S}_{\rm sh}(|x_2 - x_1|)
=
  {\cal S}_{\loarrow{\rm sh}}(x_2,x_1)
- {\cal S}_{\loarrow{\rm sh}}(x_1,x_2)
\label{shotn1}
\end{equation}

\noindent
where the unidirectional term ${\cal S}_{\loarrow{\rm sh}}(x_j,x_i)$
correlates the induced flux at $x_j$ with the random inducing
flux at $x_i$, and has a structure determined as follows.
To begin with, note that the correlated two-particle fluctuation
at time $t$, following a spontaneous change $\delta N_{s''}$
in the population of spin subband $s''$, is

\[
{\Bigl( {\widetilde R}_{\alpha \alpha'}(t)
- {\widetilde R}_{\alpha \alpha'}(\infty) \Bigr)}
\delta f_{\alpha'}
= {\widetilde C}_{\alpha \alpha'}(t) 
{ {\delta f_{\alpha'}}\over {\delta N_{s''}} } \delta N_{s''}.
\]

\noindent
Coulomb screening is fully incorporated.
When a particle is added at $x_1$ we have $\delta N_{s''} = +1$;
when a particle is removed at $x_2$, then $\delta N_{s''} = -1$.
The sign of $\delta N_{s''}$ determines the sign of
the corresponding unidirectional term in Eq. (\ref{shotn1}).

We next observe that, for each of the active carriers in the segment's
population $N = \sum_{s''} N_{s''}$,
the time of arrival at the source boundary
is uncorrelated with all other arrival times.
(In the same way, departure times at the drain are mutually uncorrelated.)
It follows that ${\cal S}_{\loarrow{\rm sh}}$
is an incoherent sum of transients:

\begin{eqnarray}
{\cal S}_{\loarrow{\rm sh}}(x_j,x_i)
\buildrel \rm def \over =&&
2 e^2 \sum_{s''} N_{s''}
{\Biggl\{
\sum_{\alpha[x = x_j]} \sum_{\alpha'[x' = x_i]}
\int^{\infty}_0 \!\!\ dt {~}
(v_x)_{\bf k} {\widetilde C}_{\alpha \alpha'}(t)
(v_x)_{\bf k'}
{{\delta f_{\alpha'}}\over {\delta N_{s''}}} |\delta N_{s''}|
\Biggr\}}
\cr
=&&
2 e^2 {N\over {\Delta N}}
\int d^{\nu} r \delta(x - x_j) \int d^{\nu} r' \delta(x' - x_i)
{1\over \gamma_{C}}
{\langle\!\langle
v_x v'_x {\widetilde \Delta} {\rm f}^{(2)} ({\bf r}, {\bf r'}; 0)
\rangle\!\rangle}_c',
\label{shotn1.1}
\end{eqnarray}

\noindent
where we have used Eq. (\ref{scrg.2}) with the spin trace

\[
\sum_{s''} N_{s''} { {\delta f_{\alpha'}}\over {\delta N_{s''}} }
= {N\over 2} \sum_{s''} { {\delta f_{\alpha'}}\over {\delta N} }
{ {\delta N}\over {\delta N_{s''}} }
= {{N {\widetilde \Delta} f_{\alpha'}}\over {{\widetilde \Delta} N}}.
\]

Our many-body construction of shot noise rests on two assumptions.
The first is ergodicity, after the original argument of Schottky:
a typical carrier in the segment must enter through the source (cathode)
and, eventually, leave through the drain (anode). There is no
temporal correlation among individual transits.
The second assumption is that each carrier in the ensemble has distinct
roles as both agent and spectator: it generates shot noise,
and it is also part of the many-body response making up
the shot noise. These dual roles are statistically independent.

The phenomenological content of Eqs. (\ref{shotn1}) and (\ref{shotn1.1})
is the same as for the Boltzmann-Langevin formalism
\cite{kogan}
except that
there is no longer any need for commitment to a specific collisional form
(other than expecting it to satisfy conservation).
Notice too that the total shot noise vanishes identically at equilibrium
because detailed balance renders the equation of motion
for the resolvent $R^{\rm eq}(t)$ collisionless, and hence self-adjoint.
Self-adjointness is preserved for ${\widetilde R}^{\rm eq}(t)$
since Coulomb forces are conservative. One can then show that

\[
{\langle\!\langle
v_x {\rm Re}{\{{\widetilde {\rm C}}^{\rm eq}_{\alpha \alpha'}(\omega)\}}
v'_x {\widetilde \Delta} f^{\rm eq}({\bf r'})
\rangle\!\rangle}'
\equiv
{\langle\!\langle
{\widetilde \Delta} f^{\rm eq}({\bf r}) v_x
{\rm Re}{\{{\widetilde {\rm C}}^{\rm eq}_{\alpha \alpha'}(-\omega)\}} v'_x
\rangle\!\rangle}'.
\]

\noindent
This produces exact cancellation between the right-hand
terms of Eq. (\ref{shotn1}).

Significantly, the Thomas-Fermi screening coefficient $\gamma_{C}$
dividing the flux autocorrelation in the second line of
Eq. (\ref{shotn1.1}), is cancelled exactly by its presence
within the autocorrelation via Eqs. (\ref{scr10}) and (\ref{scr16}).
This means that the only type of Coulomb screening
affecting the shot noise is collision-mediated.
Indeed, any homogeneous renormalisation of the equilibrium fluctuations
leaves the ratio $\Delta f/\Delta N$ untouched.
Consequently such a rescaling has absolutely no influence on the shot noise.
By comparison, the effect on the free-electron Johnson noise can be dramatic
\cite{gc}
since it is proportional to $\gamma_{C}$.

The contrast between their Coulomb responses is one
demonstration that thermal noise and shot noise are
in fact distinct many-body phenomena. Although they share
a common microscopic structure in Eq. (\ref{scr16}),
in thermodynamic terms one is an extensive continuum
quantity driven by fluctuations of the kinetic energy,
while the other is short-ranged and corpuscular,
driven by fluctuations of the local particle number.
For strongly degenerate systems the two are
disproportionate because of
the scale difference $N/{\widetilde \Delta} N$,
which becomes unity only in the classical limit.
It follows that, unlike the perturbative treatments,
\cite{theor}
this formalism will not admit a universal interpolation formula giving
thermal noise in one regime and (true) shot noise elsewhere.
\cite{analyt}

We propose a simple experimental test of incommensurability,
applicable at any current.
In a point-contact constriction defined on a
two-dimensional electron gas at a heterojunction,
thermal and shot noise are both measurable.
\cite{rez,ksgje}
Thermal noise, by its scaling with $\gamma_{C}$,
depends strongly on electron density.
\cite{gc}
Shot noise does not share this dependence.
If the density in the channel is changed,
for example by back-gate biasing,
the thermal noise should vary strongly with the bias voltage.
By contrast, the shot noise should have none of this variation
since it is immune to self-screening
of the carrier fluctuations in the quantum well.

Our direct concern is the high-field limit, where inelastic collisions rule.
For an initial look at high-field shot noise
we explore the Drude model of a uniform wire,
emphasising that its low-field behaviour,
though revealing, does not address
the elastically dominated diffusive regime.
\cite{theor}
We take the case where
the segment and its leads are physically identical.
\cite{sw0}
This is an idealised example;
experimentally there must be some differentiation
between sample and reservoirs, expressible in
known boundary conditions.

The region has length $l = x_2 - x_1 \ll L_x$, where
$L_x$ is the length of the complete assembly, segment plus leads.
Since at least one of the dimensions may approach
the mean free path, the BGFs for this problem have the short-range
spatial structure detailed in Appendix {\ref{apxSN}} and
modified by collisional Coulomb effects.
While we do not analyse the latter in computational detail here,
we propose a useful approximation based on
the Ansatz of Eq. (\ref{apxcsd3.1}) for the operator $\Gamma$
applied to Eqs. (\ref{scr10}) and (\ref{scr15}).
In a conductor of diameter much greater than the mean free path,
we take the bulk Fourier coefficients
${\rm C}^{(b)}({\bf q},0)$ 
and $\gamma_{\rm coll}({\bf q},0)$, respectively
for the correlated propagator of Eq. (\ref{apxSN7.0})
and the collision-mediated Coulomb suppression
of Eq. (\ref{apxcsd2}). The unidirectional shot noise of
Eq. (\ref{shotn1.1}) then becomes

\begin{eqnarray}
{\cal S}_{\loarrow{\rm sh}}(x_j,x_i) \approx&&
2 {{n e^2}\over {\Delta n}}
\int\!\!\!\!\int d^{\nu-1}r_{\perp}
\int\!\!\!\!\int d^{\nu-1}r'_{\perp}
\int {d^{\nu}q\over (2\pi)^{\nu}}
\exp \!{\{i[q_x(x_j - x_i) + {\bf q}_{\perp}{\bbox \cdot}
({\bf r}_{\perp} - {\bf r'}_{\perp})]\}} {~}
\cr
&&\times
\gamma^2_{\rm coll}({\bf q},0)
\int {2d^{\nu}k\over (2\pi)^{\nu}}
\int {d^{\nu}k'\over (2\pi)^{\nu}}
(v_x)_{\bf k}
{\rm C}^{(b)}_{{\bf k}{\bf k'}}({\bf q},0)
(v_x)_{\bf k'} \Delta f_{\bf k'}
\label{shotn2}
\end{eqnarray}

\noindent
where, for any wave vector ${\bf u}$, we write
its transverse component as ${\bf u}_{\perp}$.
We have dropped a contribution $\sim \gamma^3_{\rm coll}$
coming from the second term on the right-hand side
of Eq. (\ref{scr15}) for ${\rm {\widetilde C}}$.
In a full study of nonequilibrium Coulomb processes
within shot noise, $\gamma_{\rm coll}$ is clearly central.
Note that there are no transverse terms in one dimension (1D),
nor is it strictly possible to discuss
semiclassical Coulomb effects in 1D.

\section{Calculations}

In this Section we present calculations for
our inelastic model. We omit Coulomb screening, letting
$\gamma_{\rm coll} \to 1$ in Eq. (\ref{shotn2}).
To build up a detailed picture we start with shot noise in
a one-dimensional wire.

\subsection{One Dimension}

For a segment much shorter than the total
system size, we can simplify the calculation by setting
${\rm C}^{(b)}(q,0) \approx {\rm C}^{(0)}(q,0)$;
see Eqs. (\ref{apxSN5.0}) and (\ref{apxSN7.0}).
The omitted term is proportional to the finite resolution
function $\varphi_1(q; {1\over 2}L_x) \sim  \delta(q)/L_x$.
Since, over the segment,
most of the structure involves $ql \gtrsim 1$,
the approximation results in a negligible error of order $l/L_x \ll 1$.

Proceeding from Eq. (\ref{shotn2}) with $\nu = 1$ and using
Eq. (\ref{apxSN6.1b}) for ${\rm C}^{(0)}$ we obtain

\begin{eqnarray}
{\cal S}_{\loarrow{\rm sh}}(\xi)
=&&
2 {{n e^2}\over {l \Delta n}} {\left( {\hbar\over m^*} \right)}^2
\int k {dk\over \pi} \int k' dk'{\tau\over k_d}
\theta(k - k') e^{-(k - k')/k_d} {\Delta f_{k'}}
\cr
&& \times
\int l{{dq}\over {2\pi}}
\exp \!{\left[ iq{\left(
      \xi l - {{\hbar \tau}\over {2 m^* k_d}}(k^2 - {k'}^2)
   \right)}
\right]}
\cr
{\left. \right.} \cr
=&&
2 {{{n} e^2 \tau}\over {m^* l}}
{\left( {\hbar^2\over {m^* \pi \Delta n}} \right)}
{1\over {k_d}}
\int k dk e^{-k/k_d} \int^k_{-\infty} k' dk'
e^{k'/k_d} {\Delta f_{k'}}
\delta{\Bigl( \xi - (k^2 - {k'}^2)/p_d^2 \Bigl)}.
\label{apx-ish1}
\end{eqnarray}

\noindent
We have introduced $\xi = (x_j - x_i)/l$ for $i,j = 1,2$
and the wave number $p_d$ defined by
$p_d^2 = 2m^* l k_d / \hbar \tau = 2m^* eV/\hbar^2$.
Next, use the expression for $\Delta f$ in terms of $\Delta f^{\rm eq}$;
see Eq. (\ref{apxSN13a}). Rearranging the order of integration, we get

\begin{eqnarray}
{\cal S}_{\loarrow{\rm sh}}(\xi)
=&&
2 {{n e^2 \tau}\over {m^* l}}
{\left( {{\hbar^2 p_d^2}\over {2 m^* }} \right)}
\int {{dk''}\over \pi}
{{\Delta f^{\rm eq}_{k''}}\over {\Delta n}}  e^{k''/k_d}
\int^{\infty}_{k''} {{k dk}\over k_d^2} e^{-k/k_d}
\int^{|k|}_{|k''|} 2k' dk'
\delta({k'}^2 + \xi p_d^2 - k^2)
\cr
{\left. \right.} \cr
=&&2eI
\int^{\infty}_0 {{dk''}\over \pi}
{{\Delta f^{\rm eq}_{k''}}\over {\Delta n}}
{\Biggl\{
e^{k''/k_d} 
\int^{\infty}_{k''} {{kdk}\over k_d^2} e^{-k/k_d}
\theta(\xi) \theta{\Bigl( k - \sqrt{k''^2 + p_d^2} \Bigr)}
\Biggr.} \cr
&& +{~}
{\Biggl.
e^{-k''/k_d}
\int^{\infty}_{-k''}{{kdk}\over k_d^2} e^{-k/k_d}
  {\left[
   \theta(\xi) \theta{\Bigl( k - \sqrt{k''^2 + p_d^2} \Bigr)}
 - \theta(-\xi) 
   \theta{\Bigl( k'' - \sqrt{k^2 + p_d^2} \Bigr)}
  \right]}
\Biggr\}}
\label{apx-ish2}
\end{eqnarray}

\noindent
where $I = (n e^2 \tau/m^* l) V$ is the current through the wire.
We now go to the degenerate limit, replacing
the equilibrium fluctuation according to
$\Delta f^{\rm eq}_k = (m^* k_BT/\hbar^2 k_F)\delta (|k| - k_F)$.
The number-fluctuation density is $\Delta n = 2m^*k_BT/\hbar^2\pi k_F$.
A series of manipulations leads to the total shot noise

\begin{eqnarray}
{\cal S}_{\rm sh}(l)
=&&
{\cal S}_{\loarrow{\rm sh}}(+1) - {\cal S}_{\loarrow{\rm sh}}(-1)
\cr
{\left. \right.} \cr
=&&
2eI
{\left\{
e^{-\sqrt{k_F^2 + p_d^2}/k_d}
{\left[ 1 + {{\sqrt{k_F^2 + p_d^2}}\over k_d} \right]}
\cosh \!{\left( {k_F\over k_d} \right)}
- \theta(\varepsilon_F - eV) e^{-k_F/k_d}
\right.}
\cr
{\left. \right.} \cr
&& \times
{\left.
   {\left[
   {{\sqrt{k_F^2 - p_d^2}}\over k_d}
      \cosh \! {\left( {{\sqrt{k_F^2 - p_d^2}}\over k_d} \right)}
    - \sinh \! {\left( {{\sqrt{k_F^2 - p_d^2}}\over k_d} \right)}
  \right]}
\right\}}.
\label{apx-ish4}
\end{eqnarray}

\noindent
Again we see the non-analyticity of this expression with respect
to the driving field. In this uniformly-embedded wire model,
expansion of the shot noise
in powers of $V$ is not valid at low fields.
On the contrary, the shot noise becomes perturbative in $1/V$ at
high fields, which could be termed the Schottky domain.
The expression is perfectly calculable
and there are two asymptotic cases of interest.

(a) High fields, $eV \gg \varepsilon_F$:

\begin{eqnarray}
{\cal S}_{\rm sh}(l)\Big|_{V \to \infty}
=&& 2eI {\left( 1 + {p_d\over k_d} \right)} e^{-p_d/k_d}
\cr
{\left. \right.} \cr
\to&& 2eI {\left( 1 - {{m^*(l/\tau)^2}\over {eV}} \right)}.
\label{apx-ish7}
\end{eqnarray}

\noindent
The high-field limit gives the full Schottky expression
$2eI$ with a correction, dependent on the wire parameters,
which is asymptotically negligible.
The same formal result also holds for any chosen value of $I$
in the collisionless regime $\tau \to \infty$,
confirming that our model recovers the shot-noise behaviour
of a monoenergetic flux.

(b) Low fields, $eV \ll \varepsilon_F$.
The mean free path is $\lambda = \tau v_F$. Then
$(k_F^2 \pm p_d^2)^{1\over 2} \to
k_F [1 \pm lk_d/\lambda k_F  - {1\over 2}(lk_d/\lambda k_F)^2]$ and

\begin{equation}
{\cal S}_{\rm sh}(l)\Big|_{V \to 0}
= 2eI {\left( 1 + {l\over \lambda} + {l^2\over 2\lambda^2} 
\right)} e^{-l/\lambda}.
\label{apx-ish6}
\end{equation}

\noindent
Our 1D inelastic model again gives $2eI$ at low fields,
with no suppression in the ballistic limit $\lambda \gg l$.
There is, hovever, exponential decay of the shot noise as the
wire length increases beyond the mean free path. The result is
understandable as source and drain currents rapidly
decorrelate with increasing $l/\lambda$;
if $\lambda \propto \tau$ is made smaller while $l$ is kept fixed,
the exponential attenuation is broadly consistent
with Monte Carlo results of Liu, Eastman, and Yamamoto.
\cite{yam}
Their more general simulation of 1D low-field shot noise,
which includes both elastic scattering and inelastic phonon emission,
exhibits strong suppression in the inelastically dominated regime.

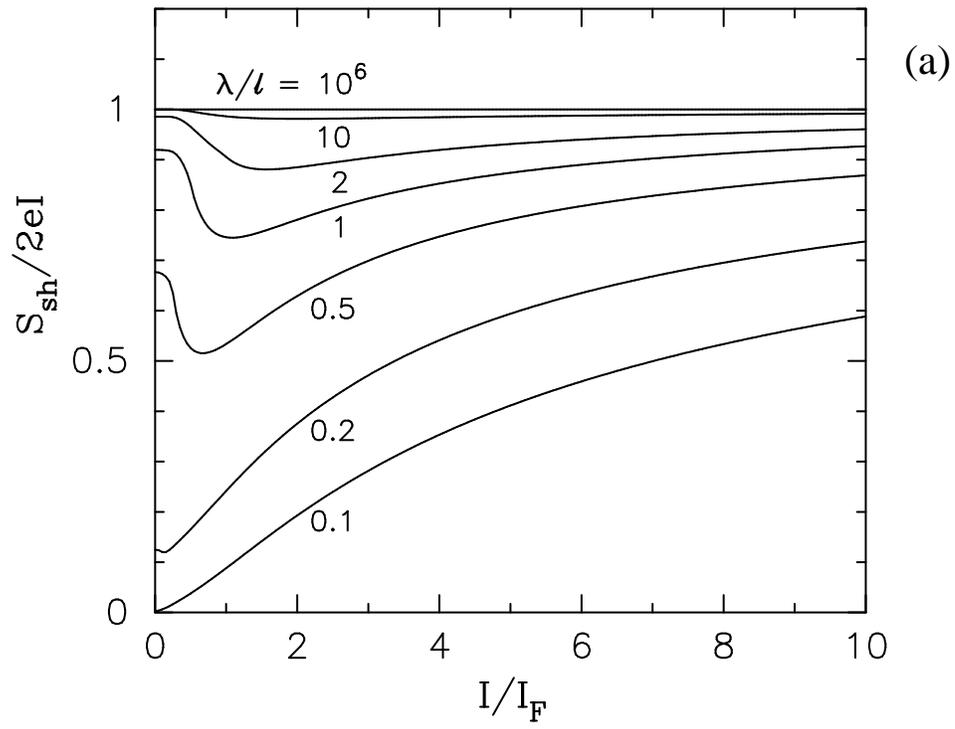
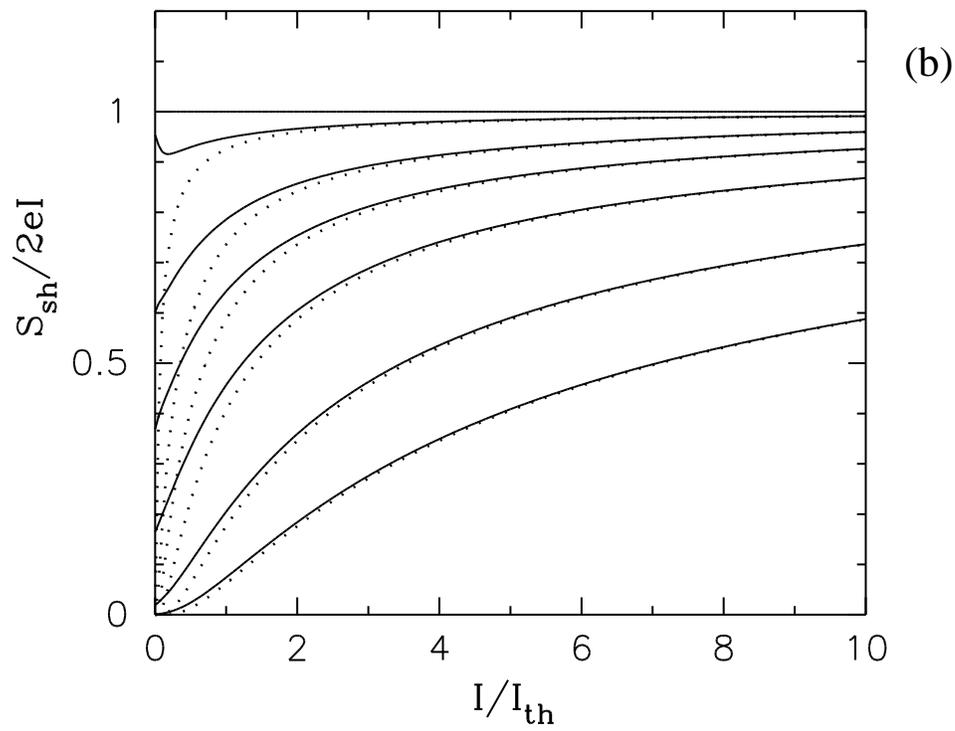
\begin{figure}
\caption{
(a)
Shot noise in the Drude model of a degenerate one-dimensional wire,
as a function of current in units of the Fermi current
and normalised to full shot noise. Each curve is for a fixed
ratio $\lambda/l$ of mean free path to length of wire;
the same set of ratios is used for all subsequent figures.
At high currents and in the ballistic limit (upper curves),
the shot noise tends to its full value.
At low currents and away from the ballistic limit, degeneracy
inhibits the natural tendency of the shot noise to
exponential suppression with increased sample length.
(b) Full line: shot noise of classical carriers, as a function
of current in units of the thermal current. The attenuation
at low currents is much more pronounced than in (a). Dots:
high-field asymptote defined by Eq. (\ref{apx-ish7}). Note how
both classical and fermionic results rapidly assume this form at
higher currents.
}
\label{fig2}
\end{figure}

In Fig. \ref{fig2} we compare the 1D shot noise
normalised to $2eI$, for degenerate and
classical conductors. Fig. \ref{fig2}(a) shows the results for a
degenerate sample, as a function of current normalised
to $I_F = nev_F$. The plots are for a range of wire lengths from
the ballistic limit $l = 10^{-6}\lambda$, up to $l = 10\lambda$.
At low fields the intercepts at $I = 0$, given by Eq. (\ref{apx-ish6}),
show their attenuation away from
the ballistic limit. For higher currents, the shot noise
quickly settles to the form given by Eq. (\ref{apx-ish7}).

In Fig. \ref{fig2}(b) we plot, for comparison, the shot noise of a 1D
system whose carrier distribution is classical:
$\Delta f^{\rm eq}_k = f^{\rm eq}_k \propto
\exp(-\varepsilon_k/m^* v_{\rm th}^2)$
where $v_{\rm th} = (k_BT/m^*)^{1\over 2}$.
The current is normalised to $I_{\rm th} = ne\sqrt{2}v_{\rm th}$
and $\lambda = v_{\rm th} \tau$..
Also plotted is the asymptotic
form appearing in the first line of Eq. (\ref{apx-ish7}).
Again at low currents we see attenuation
with increasing wire length, stronger than in Fig. \ref{fig2}(a).
At higher currents there is the same rapid convergence
to the asymptotic result (evident at surprisingly modest currents)
as found in Fig. \ref{fig2}(a).

From our comparison of Fermi-Dirac and Maxwell-Boltzmann
versions of the model, we conclude that degeneracy
contributes mainly at currents below $I_F$. For
$eV < \varepsilon_F$ the driving voltage cannot
overcome the collective stability of the Fermi sea,
and the zero-current correlations persist.
For $eV \gtrsim \varepsilon_F$ appreciable redistribution of the
particle occupancies suddenly becomes possible,
with an initial dip in relative correlation strength. At higher
fields most electrons move independently and ballistically,
in the sense that $\tau {\langle v f \rangle}/n \gg l$.
The shot noise is then in the Schottky domain.

\subsection{Two and Three Dimensions}

For higher dimensions we must include traces over the transverse
degrees of freedom.
Write $A \equiv 2R$ for $\nu = 2$ and $A \equiv \pi R^2$
for $\nu = 3$ where $R$ is the half-width (for a strip) or the radius
(for a cylinder).
As in the 1D case, the condition $l \ll L_x$
implies that the correlated BGF is well approximated by

\[
{\rm C}_{{\bf k} {\bf k'}}({\bf q}, {\bf q'}, \omega)
= A \delta(q_x - q'_x)
\varphi_{\nu - 1}({\bf q}_{\perp} - {\bf q'}_{\perp}; R)
{\rm C}^{(0)}_{{\bf k} {\bf k'}}({\bf q}, \omega);
\]

\noindent
refer to Appendix {\ref{apxSN}} for details.
After integrating over the cross-sectional co-ordinates
and applying Eq. (\ref{apxSN6.1b}) for ${\rm C}^{(0)}$,
Eq. (\ref{shotn1.1}) reads

\begin{eqnarray}
{\cal S}_{\loarrow{\rm sh}}(\xi)
=&&
2 {{n e^2 A }\over {l \Delta n}}
{\left( {\hbar\over m^*} \right)}^2
\int {{2k_xdk_x}\over (2\pi)^{\nu}}
\int k_x' dk_x'{\tau\over k_d}
\theta(k_x - k_x') e^{-(k_x - k'_x)/k_d}
\cr
&& \times
\int d^{\nu-1}k_{\perp} {\Delta f_{{\bf k'}}}
F_{\nu}(a)
\int l{{dq}\over {2\pi}}
\exp \!{\Bigl[
ilq {\Bigl( \xi - p_d^{-2}(k_x^2 - {k'}_x^2) \Bigr)}
\Bigr]}.
\label{apx-ish8}
\end{eqnarray}

\noindent
The shape factor $F_{\nu}(a)$, whose argument is
$a = \hbar \tau |k_x - k'_x| k_{\perp}/(2m^* k_d R)$, has the form

\begin{mathletters}
\label{apx-ishff}
\begin{eqnarray}
F_{\nu}(a)
=&& A^2
\int { {d^{\nu-1} q_{\perp}}\over {(2\pi)^{\nu-1}} }
\varphi_{\nu-1}({\bf q}_{\perp}; R)
\exp \!{\left(
  -{{i\hbar \tau (k_x - k'_x)}\over {m^* k_d}}
   {\bf k}_{\perp}{\bbox \cdot}{\bf q}_{\perp}
    \right)}
\cr
&& \times
\int { {d^{\nu-1} q'_{\perp}}\over {(2\pi)^{\nu-1}} }
\varphi_{\nu-1}({\bf q}_{\perp} - {\bf q'}_{\perp}; R)
\varphi_{\nu-1}({\bf q'}_{\perp}; R)
\label{apx-ishffa},
\end{eqnarray}

\noindent
which reduces to the expressions

\begin{equation}
F_2(a) = \theta(1 - a) {\Bigl( 1 - a \Bigr)},
\label{apx-ishffb}
\end{equation}

\begin{equation}
F_3(a) = \theta(1 - a)
{\left[
1 -  {2\over \pi}{\left( \arcsin a + a \sqrt{1 - a^2} \right)}
\right]}.
\label{apx-ishffc}
\end{equation}
\end{mathletters}

\noindent
In a wire of finite width, this function directly expresses
the constraint on lateral motion of the carriers;
it cross-couples, kinematically, the
transverse and longitudinal modes. Here there is none
of the dynamical cross-coupling induced, for example,
by elastic scattering.
Kinematic suppression is inherent in the form of ${\rm C}^{(0)}$
[Eqs. (\ref{apxSN6}) and (\ref{apxSN6.1})], itself
conditioned by the free-streaming operator in the Boltzmann equation.
It is not surprising to find an entirely geometric
source of shot-noise suppression in
two and three dimensions (2D; 3D).
This echoes, in part, Landauer's remark
\cite{landauer2}
on the need to sample more than just the longitudinal
trajectories in any semiclassical calculation.

We process Eq. (\ref{apx-ish8}) along lines analogous to
Eq. (\ref{apx-ish2}), going to the degenerate limit
with Eq. (\ref{apxSN13b}) for $\Delta f$
and the fluctuation density
$\Delta n = m^*k_BTk_F^{\nu-2}/\hbar^2 \pi^{\nu -1}$.
The first shot-noise component reduces to

\begin{mathletters}
\label{apx-ish13}
\begin{eqnarray}
{\cal S}_{\loarrow{\rm sh}}(+1)
=&&
2eI
\int^{\infty}_{p_d} {{k_x dk_x}\over k_d^2} e^{-k_x/k_d}
\int^{k_F}_0 {\left( {{2k_F}\over {\pi k_{\perp}} } \right)}^{3 - \nu}
{dp_{\perp}\over 2k_F}
\cr
&& \times
{\Biggl[
2\theta(p_x - p_{\perp}) \cosh \!{\left( {p_{\perp}\over k_d} \right)}
F_{\nu}(a_-)
\Biggr.} \cr
&&
{\Biggl.
+ \theta(p_{\perp} - p_x) e^{-p_{\perp}/k_d}
{\Bigl( F_{\nu}(a_-) - F_{\nu}(a_+) \Bigr)}
\Biggr]}
\label{apx-ish13a}
\end{eqnarray}

\noindent
where $p_{\perp} = (k_F^2 - k_{\perp}^2)^{1\over 2}$ and
$p_x = (k_x^2 - p_d^2)^{1\over 2}$;
the shape-factor arguments are
$a_{\pm} = \hbar \tau (k_x \pm p_x) k_{\perp}/(2m^* k_d R)$.
Similarly the second component is

\begin{eqnarray}
{\cal S}_{\loarrow{\rm sh}}(-1)
=&&
2eI \theta(\varepsilon_F - eV)
\int^{k_F}_{p_d} {{k_x dk_x}\over k_d^2}
\int^{k_F}_{k_x}
{\left( {{2k_F}\over {\pi k_{\perp}} } \right)}^{3 - \nu}
{dp_{\perp}\over 2k_F} e^{-p_{\perp}/k_d}
\cr
&& \times
{\Biggl[
e^{p_x/k_d} F_{\nu}(a_-) - e^{-p_x/k_d} F_{\nu}(a_+)
\Biggr]}.
\label{apx-ish13b}
\end{eqnarray}
\end{mathletters}

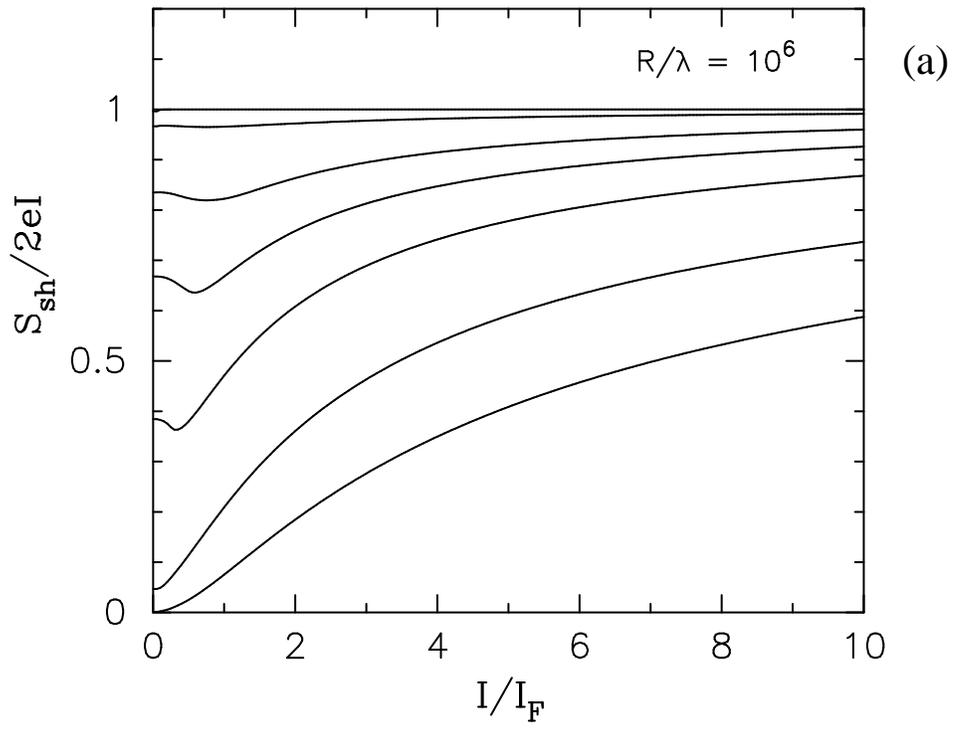
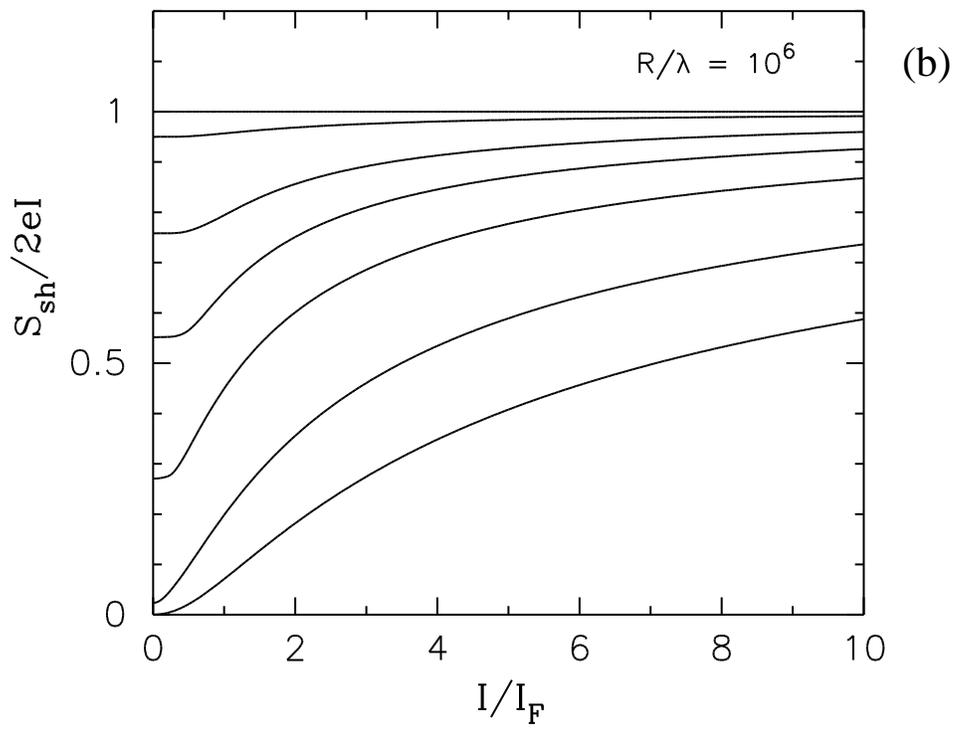
\begin{figure}
\caption{
Shot noise in the degenerate Drude model for (a) very wide
two-dimensional strips, and (b) three-dimensional cylinders
of very large radius. The behaviour at low currents differs
significantly from Fig. \ref{fig2}(a), with progressively greater
suppression at longer wire lengths and higher dimensionality.
At high currents the asymptotic behaviour is identical with that
in one dimension.
}
\label{fig3}
\end{figure}

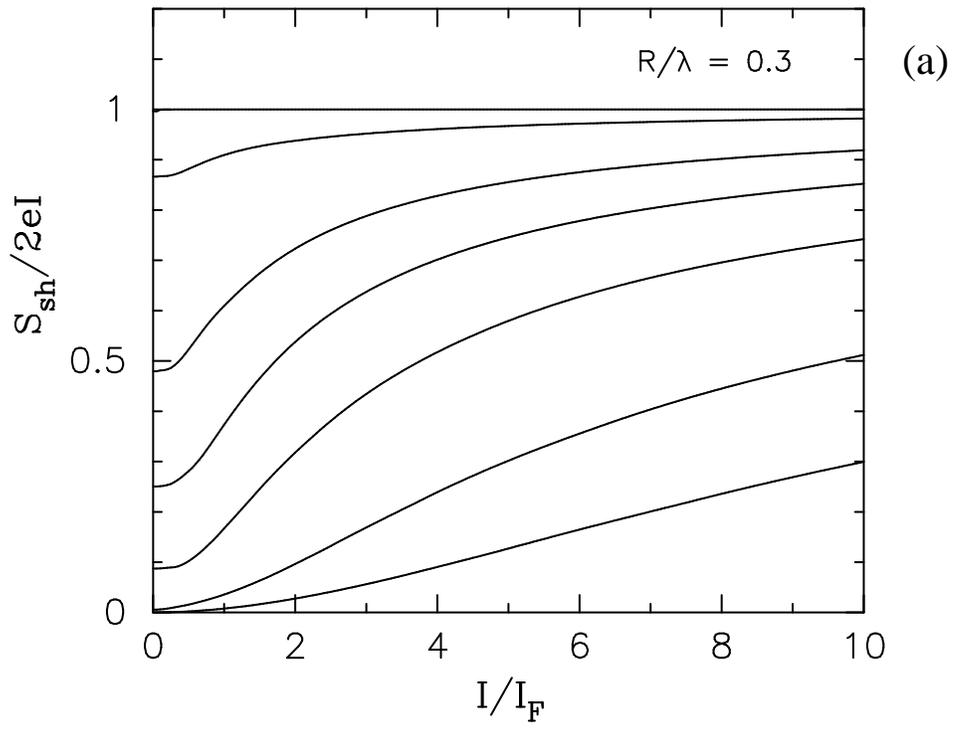
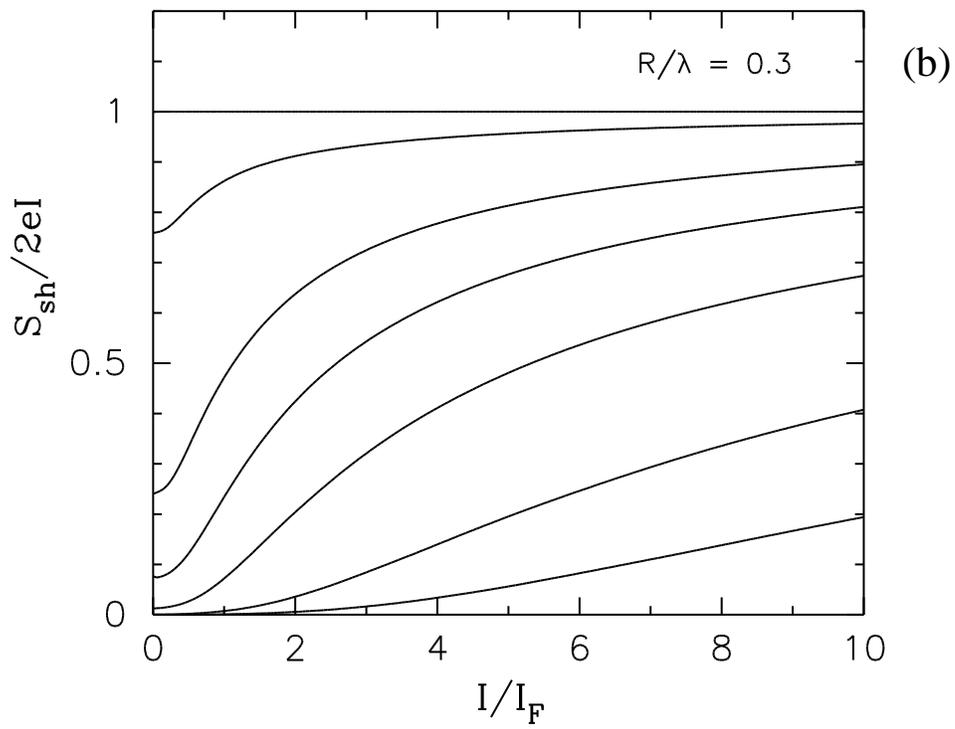
\begin{figure}
\caption{
Shot noise in the degenerate Drude model for (a) narrow
strips, and (b) thin cylinders.
Away from the ballistic limit (topmost curves) there
is remarkable shot-noise suppression over the entire range of the
current. This effect is inherent in the kinematic term of
the higher-dimensional Boltzmann equation; it cannot be
simulated by a one-dimensional approximation.
}
\label{fig4}
\end{figure}

\noindent
In Figs. \ref{fig3} and \ref{fig4} we plot the shot noise
in two and three dimensions
for a range of wire geometries, as a function of current normalised to
$I_F = neAv_F$. The results in Fig. \ref{fig3} are calculated for the
thick-wire limit $R \gg \lambda$;
those in Fig. \ref{fig4} are for a thin wire, $R = 0.3\lambda$.
The same values of $\lambda/l$ are used as in Fig. \ref{fig2},
running down monotonically from the top.

For both strips and cylinders in Fig. \ref{fig3},
it is the behaviour of ${\cal S}_{\rm sh}$ at higher values of $I$
that comes to notice first: each curve merges with its asymptotic
1D analogue illustrated in Fig. \ref{fig2}(b).
As will shortly become clear,
it is only in the thick-wire limit, and {\em then} only for high currents,
that a 1D treatment can in any sense mimic
the exact calculations for higher dimensions
[those conditions amount to setting $F_{\nu} \approx 1$
within Eq. (\ref{apx-ish13})].
In fact, as one progresses from 1D through 3D
[Figs. \ref{fig2}(a), \ref{fig3}(a), and \ref{fig3}(b)],
the zero-current intercepts of the curves
indexed by the same $\lambda/l$
undergo a marked and systematic increase in suppression.
This shows (at least in the simple Drude model)
that the 1D calculation is a poor estimate of low-field noise
in realistic geometries, even in the thick-wire limit.

We come to the thin wires of Fig. \ref{fig4}.
The uppermost curves, in the ballistic regime $\lambda \gg l$,
are unchanged from Figs. \ref{fig2} and \ref{fig3}.
However, relative to Fig. \ref{fig3}
the remaining curves in Fig. \ref{fig4} change dramatically
as one moves further from the ballistic limit.
There is now substantial shot-noise suppression
throughout the whole range of $I$.
For example, in Fig. \ref{fig3}(b)
the longest 3D wire, $l = 10\lambda$, has
${\cal S}_{\rm sh}/2eI = 0.59$ at the highest current $I = 10I_F$,
while its opposite number in Fig. \ref{fig4}(b)
reaches the value 0.2; a threefold reduction.
Calculations at much higher fields confirm the eventual
recovery of full shot noise in keeping with Eq. (\ref{apx-ish7}).

The outcome of spatially constrained carrier motion is the
extensive suppression of shot noise over a wide range of the current.
We stress that the effect is implicit in the generic structure
of the Boltzmann equation, and its propagators, for 2D and 3D
[see Eq. (\ref{apxSN6})]; it is simply absent in 1D.
Therefore, this mode of suppression
cannot be simulated by any one-dimensional scheme.

In a 3D wire the Fermi wavelength is 0.05 nm
at metallic electron densities.
If $\lambda = 50 {\rm ~nm}$, typical for strong inelastic scattering,
a wire of width $\sim 30 {\rm ~nm}$ would exhibit kinematic shot-noise
suppression at large currents, providing that it was not
masked by collisional Coulomb effects.
In future we plan to assess the latter quantitatively;
the comparative action of nonequilibrium screening,
in 2D versus 3D, should itself be an interesting window
on how dimensionality affects fluctuations.
\cite{reggi2}

\section{Summary}

We have described and implemented a nonperturbative microscopic formalism
for current fluctuations in metallic systems, down to the mesoscopic scale,
within the ambit of semiclassical theory. Our strategy for incorporating
fermion correlations into the Boltzmann picture safeguards the conservation
laws at both the single-particle level and at the level of dynamic
two-particle processes, the key to nonequilibrium current noise.
In particular we have derived the nonlinear analogue of the
fluctuation-dissipation theorem. It should also be straightforward
to include semiclassical electron-electron scattering in our description
of transport and noise.

Our formalism's calculability stems from the flexible structure of the
Boltzmann-Green functions. These serve as semiclassical propagators of the
electronic Fermi-liquid correlations, mapping them uniquely to the
correlations of the nonequilibrium system.
We have demonstrated their usefulness in shedding light on the physics
of high-current shot noise, and on the importance of treating dimensionality
correctly in constricted mesoscopic samples.

The present account of nonequilibrium fluctuations raises a variety of
interesting questions, practical and abstract.
By far the most salient is the relation
between thermal noise and shot noise; our claim that the two are
thermodynamically incommensurate should be easy to test.
Thermal-noise measurements on a gated two-dimensional mesoscopic wire,
defined on a III-V heterojunction,
ought to give a strong
gate-voltage-dependent signature of Thomas-Fermi suppression from
self-confinement of the carriers in their quantum well.
Measurements of the shot noise in the same structure
should give no such signature.

We end with just two out of many theoretical issues. The first is the
role of non-analyticity of the BGF solutions at
low fields, and the implications for semiclassical linear response.
The phenomenon is well known in uniform systems,
\cite{sw0,bakshi}
where the free-streaming part
of the Boltzmann operator is manifestly anomalous in its
vanishing with the applied field.
While we have no firm information on whether the BTE for nonuniform systems
shares this behaviour, we point out that our shot-noise results, obtained
in a spatially inhomogeneous model (albeit weakly so), are certainly
non-analytic in the applied voltage. The low-field asymptotics of the
general Boltzmann equation  have an obvious bearing on how semiclassical
noise is to be calculated, and their clarification would be a significant
advance.

Second, there is the  status of the BGF approach within quantum kinetics.
In terms of, say, the Kadanoff-Baym analysis of the Boltzmann equation,
\cite{BK}
our semiclassical equation of motion for the fluctuations should emerge from
the quantum evolution of the particle-hole amplitudes
in the long-wavelength limit,
\cite{kogan2}
much as the ordinary BTE is distilled from the long-wavelength
dynamics of the density matrix for a single particle.

\section*{Acknowledgments}

We are indebted to the late R. Landauer for his encouragement,
which stimulated the present work in its earliest stages.
We thank E. Davies for help with calculations and figures
and N. W. Ashcroft, R. J-M. Grognard, and D. Neilson for
fruitful discussions.

Last, but not least, we acknowledge the cultural contribution of
Physical Review B and its editors. But for that journal's
egregious year-long dissembling over its submission,
this work might still be written in American English.

\appendix

\section{Types of Screening in Nonequilibrium Systems}
\label{apxcsd}

We begin this Appendix with some basic properties of
collision-mediated Coulomb screening, which influences both
thermal and shot noise. We end with Thomas-Fermi screening,
which influences thermal noise alone.
Keeping in mind the high-field application, we base
our calculations on the collision time model
(see the following Appendices).
Our results are specific to a bulk system with
inelastic scattering; we stress both their illustrative intent
and their inapplicability to the elastic diffusive regime,
where the Boltzmann propagators and the
collisional screening effects are very different.

We take the Fourier transform of the collision-mediated
Coulomb operator [see Eq. (\ref{scr10.1})] using the
form of the correlated BGF in the bulk limit,
Eq. (\ref{apxSN7.0}).
The (spin-independent) function ${\Gamma}$ is

\begin{equation}
{\Gamma}_{{\bf k} {\bf k'}} ({\bf q}; \omega) =
\Omega \delta_{{\bf k} {\bf k'}}
- {{2e}\over {\hbar \Omega^2}} \sum_{\bf k''}
{\rm C}^{(b)}_{{\bf k} {\bf k''}}({\bf q}; \omega)
{ {\partial f_{\bf k''}}
\over {\partial {\bf k''}} } {\bbox \cdot}
{\bbox {\cal E}}_{C}({\bf q})
\sum_{\bf k'''}
{\Gamma}_{{\bf k'''} {\bf k'}}({\bf q}; \omega),
\label{apxcsd1}
\end{equation}

\noindent
where the electron field is
$e{\bbox {\cal E}}_{C}({\bf q}) = -i{\bf q}{\rm V}_{C}(q)$
and ${\rm V}_{C}(q)$ is the Coulomb potential transform.
In three dimensions there is a complication owing
to the long-range Coulomb tail; by assumption, all fields
are shorted out beyond the system boundaries.
We model this constraint by introducing a cutoff,
so that ${\rm V}_{C}(q) = 4\pi e^2/\epsilon (q^2 + \kappa^2)$
where $\kappa^{-1} \gtrsim \Omega^{1\over 3}$ represents the
characteristic length scale beyond which the fields are zero.

The trace
$\gamma_{\rm coll}({\bf q}, \omega) \equiv
{\langle {\Gamma}({\bf q}, \omega; {\bf k'}) \rangle}$
is independent of wave vector ${\bf k'}$. This
follows from the decoupling of the internal summations over
kinematic variables in Eq. (\ref{apxcsd1}),
since the Coulomb field depends only on ${\bf q}$.
Summing both sides of Eq. (\ref{apxcsd1}) we get

\begin{eqnarray}
\gamma_{\rm coll}({\bf q}, \omega)
=&&
1 - {2e\over {\hbar \Omega^2}}
\sum_{\bf k} \sum_{\bf k'}
{\rm C}^{(b)}_{{\bf k} {\bf k'}}({\bf q}, \omega)
{ {\partial f_{\bf k'}}
\over {\partial {\bf k'}} } {\bbox \cdot}
{\bbox {\cal E}}_{C}({\bf q})
\gamma_{\rm coll}({\bf q}, \omega)
\cr
{\left. \right.} \cr
=&&
{\left[
1 - {{2i{\rm V}_{C}(q)}\over {\hbar \Omega}}
    \sum_{\bf k'}
    {\langle {\rm C}^{(b)}({\bf q}, \omega; {\bf k'}) \rangle}
    {\bf q} {\bbox \cdot}
    { {\partial f_{\bf k'}}\over {\partial {\bf k'}} }
\right]}^{-1}.
\label{apxcsd2}
\end{eqnarray}

\noindent
There is now a closed form for the Coulomb operator:

\begin{equation}
{\Gamma}_{{\bf k} {\bf k'}}({\bf q}, \omega)
= \Omega \delta_{{\bf k} {\bf k'}}
+
i{{8 \pi e^2 \gamma_{\rm coll}({\bf q}, \omega)}
    \over {\epsilon \hbar \Omega (q^2 + \kappa^2)}}
\sum_{\bf k''}
{\rm C}^{(b)}_{{\bf k} {\bf k''}}({\bf q}, \omega)
{\bf q} {\bbox \cdot}
{ {\partial f_{\bf k''}} \over {\partial {\bf k''}} },
\label{apxcsd3}
\end{equation}

\noindent
suggesting a possible approximation for the convolution
of ${\Gamma}$ with a typical distribution $F$, namely

\begin{equation}
{1\over \Omega} \sum_{\bf k'}
{\Gamma}_{{\bf k} {\bf k'}}({\bf q}, \omega) F_{\bf k'}
\approx \gamma_{\rm coll}({\bf q}, \omega) F_{\bf k}.
\label{apxcsd3.1}
\end{equation}

\noindent
This is exact in the $q \to 0$ limit and also reproduces the
exact relation
$\langle\!\langle {\Gamma}({\bf q}, \omega) F' \rangle\!\rangle'
= \gamma_{\rm coll}({\bf q}, \omega) \langle F \rangle$.
The Ansatz, which amounts to the decoupling
${\rm C}^{(b)}_{{\bf k} {\bf k'}}/
{\langle {\rm C}^{(b)}({\bf k'}) \rangle}
\sim F_{\bf k}/{\langle F \rangle}$,
tends to wash out the sharp features
of the integrand in Eq. (\ref{apxcsd3}).
 
We evaluate Eq. (\ref{apxcsd2}) for the Drude model
in the zero-field limit. Using Eq. (\ref{apxSN8}),
the trace of the correlated BGF over its leading wave vector is

\begin{eqnarray}
{\langle {\rm C}^{(b)}({\bf q}, 0; {\bf k'}) \rangle} \Big|_{E \to 0}
=&&
{1 \over
 { {\displaystyle {{i\hbar}\over m^*} }
 {\bf q}{\bbox \cdot}{\bf k'} + \tau^{-1}} }
- {{\varphi_3({\bf q})}\over {{1\over 2}{n} \Omega}}
\sum_{\bf k}
{f^{\rm eq}_{\bf k}\over
 { {\displaystyle {{i\hbar}\over m^*} }
 {\bf q}{\bbox \cdot}{\bf k} + \tau^{-1} } }.
\label{apxcsd4}
\end{eqnarray}

\noindent
There is no contribution to Eq. (\ref{apxcsd2}) from the second
right-hand term of Eq. (\ref{apxcsd4}), and so

\begin{equation}
\gamma_{\rm coll}(q, 0) \Big|_{E \to 0}
= {\left[
1 + {q^2_{\rm TF}\over {q^2 + \kappa^2}}
    {\left( 1 - {\arctan(\lambda q)\over {\lambda q}} \right)}
\right]}^{-1},
\label{apxcsd5}
\end{equation}

\noindent
where the Thomas-Fermi wave vector is defined by
$q^2_{\rm TF} = 4\pi e^2 {\Delta n}/\epsilon k_BT
= 4 e^2 m^* k_F/\pi \epsilon \hbar^2$,
and $\lambda = \tau v_F$ is the mean free path.
In the small-$q$ limit the suppression factor becomes

\begin{equation}
\gamma_{\rm coll}(q, 0) \Big|_{E \to 0}
= { {\kappa^2 + q^2}
    \over{\kappa^2 + (1 + 4\pi \sigma \tau/\epsilon)q^2} } \to 1
\label{apxcsd6}
\end{equation}

\noindent
($\sigma$ is the conductivity),
showing that there is no long-range suppression from collision-mediated
screening if the asymptotic state of the leads pins
the electric fields to zero there.
Put differently, microscopic scattering preserves
global neutrality; Thomas-Fermi screening may not,
since it redistributes charge.
This calls for the inclusion of buffer zones at the system boundaries
(in practice, several units of $q_{\rm TF}^{-1}$) to ensure
that the fields beyond the system remain evanescent.

Mesoscopically, collisional screening should be significant.
In a metal $\gamma_{\rm coll}(q,0)$
can be very small; for example, in silver at 77 K its minimum is
roughly $10^{-7}$ and its value does not rise to one half until
$q^{-1} = q_{\rm TF}^{-1} = 0.06$ nm, that is,
far below any mean free path and out of the semiclassical domain.
In heavily doped GaAs with
carrier density $n = 10^{18} {\rm ~cm}^{-3}$, comparable figures
are $\gamma_{\rm coll} = 0.01$ for the
minimum and $\gamma_{\rm coll} = 0.5$ at $q^{-1} = 10$ nm.
While these are guideline figures
for a simple model in its approach to equilibrium,
they hint at a strong role for collisional screening suppression
in high-field shot noise.

We return to Thomas-Fermi screening, associated with the
contact potentials. This is a primary source of thermal-noise suppression,
a thermodynamic effect free of any collision processes.
We outline its behaviour in a bulk jellium
conductor contacted by leads made of different jellium.
The combined system is treated as an electron gas
closed with respect to
carrier exchange, and satisfying periodic boundary conditions.
\cite{nozpin}
We must also take explicit account of the reservoir's
electrostatic potential $u_r$.

If the subsystems are macroscopic,
a term such as the second one in
the right-hand-side sum of Eq. (\ref{scruc}) goes to its
asymptotic mean $u_{c}$; the third term is negligible
to the same order. Including the explicit offset
from $u_r$, Eq. (\ref{scru1}) generalises to

\begin{eqnarray}
\gamma_{C}(q)
=&& 1 - {{\delta }\over {\delta \mu}}( u_{c} - u_r)
\cr
{\left. \right.} \cr
=&& 1
- { {({\rm V}_{C}(q) n + u_{c}) \Omega {\widetilde {\Delta}} {n}(q)}
\over {N k_BT}}
+ { {({\rm V}'_{C}(q) n' + u_r) \Omega {\widetilde {\Delta}} {n}'(q)}
\over {N' k_BT}}
\cr
{\left. \right.} \cr
=&&
1 - {\left( 1 + {u_{c}\over n {\rm V}_{C}(q)} \right)}
{\rm V}_{C}(q)
{ {\gamma_{C}(q) {\Delta n}(q)}\over k_BT }
+ {\left( 1 + {u_r\over n' {\rm V}'_{C}(q)} \right)}
{\rm V}'_{C}(q)
{ {\gamma'_{C}(q) {\Delta n}'(q)}\over k_BT },
\label{apxcsd8}
\end{eqnarray}

\noindent
where all primed quantities refer to the reservoir.
A complementary relation holds for $\gamma'_{C}(q)$.
Consider the conductor. In the long-wavelength limit we have

\[
{\rm V}_{C}(q) { {\Delta n(q)}\over k_BT }
= { q^2_{\rm TF}\over q^2}.
\]

\noindent
Putting this into Eq. (\ref{apxcsd8}) together with its
reservoir counterpart, and taking $q \ll q_{\rm TF}, q'_{\rm TF}$,
we obtain the coupled equations

\begin{equation}
{\left( q^2_{\rm TF} + {{u_{c}\Delta n}\over {k_BT n}}q^2 \right)}
\gamma_{C}(q) -
{\left( q'^2_{\rm TF} + {{u_r\Delta n'}\over {k_BT n'}}q^2 \right)}
\gamma'_{C}(q)
= q^2; {~~~~~~}
\gamma_{C}(q) + \gamma'_{C}(q) = 2,
\label{apxcsd10}
\end{equation}

\noindent
giving the limiting solutions

\begin{equation}
\gamma_{C} = 1 -
{ {q^2_{\rm TF} - {q'}^2_{\rm TF}}\over
  {q^2_{\rm TF} + {q'}^2_{\rm TF}} } {~~~} {\rm and} {~~~}
\gamma'_{C} = 1 +
{ {q^2_{\rm TF} - {q'}^2_{\rm TF}}\over
  {q^2_{\rm TF} + {q'}^2_{\rm TF}} }.
\label{apxcsd11}
\end{equation}

\noindent
In this system, a conductor that is
more metallic than the reservoir has $q_{\rm TF} > q'_{\rm TF}$,
leading to $\gamma_{C} < 1 < \gamma'_{C} < 2$.
Its thermal noise (were it accessible from outside)
would thus undergo suppression. Conversely, a relatively less
metallic sample would display {\em enhanced} thermal noise.
If $q_{\rm TF} \ll q'_{\rm TF}$, as in a lightly doped
bulk semiconductor in contact with a metal,
then $\gamma_{C}$ approaches its maximum of two.

A closed model provides an unrealistic picture of actual transport;
it may be indicative, but by no means definitive. In particular,
for truly open reservoirs, the potential $u_r$ cannot be accessed
in isolation from the chemical potential as a whole.
This means that the reciprocity between $\gamma_{C}$ and $\gamma'_{C}$,
characteristic of the closed model, does not hold. Instead, the identity
$\gamma'_{C} \equiv 1$ is a necessary boundary constraint.

For an open mesoscopic conductor, details of the contact-potential effects
on thermal noise will be sensitive both to the topology of its boundary
conditions, and to its internal electronic structure. The action of
suppression (or indeed of enhancement) is a problem for further study.

\section{Uniform Drude Model}
\label{apxT}

We derive the dynamical fluctuation structure
for a single parabolic conduction band
with uniform electron density ${n}$ and
constant mobility $\mu_e = e \tau / m^*$, where $\tau$ is the
spin-independent collision time and $m^*$ the effective mass.
The system is driven by a uniform field 
${\bf {\widetilde E}} = {\bf E} = -E {\bf {\hat x}}$
acting in the negative (drain to source) direction.
We take variations which are homogeneous over the sample region,
so that the fluctuations of interest have no spatial dependence.

The Boltzmann equation in the model is

\begin{equation}
\left[ { {\partial }\over {\partial t} } 
+  {
   { { eE}\over {\hbar} }
  }
   { {\partial }\over {\partial k_x} }
+ { 1\over \tau} \right] f_{\bf k}(t)
= 
{ {\langle f(t) \rangle} \over {\langle f^{\rm eq} \rangle} }
{ f_k^{\rm eq}\over \tau}.
\label{apxT1}
\end{equation}

\noindent
Since the Boltzmann operator is linear, the
fluctuation structure is qualitatively similar to that for
elastic scattering [differences
arise from the inhomogeneous term in $f^{\rm eq}$,
notably in the behaviours of $R(t)$ and $\Delta f(t)$].
We solve Eq. (\ref{apxT1})
by Fourier transforms in reciprocal space, so that the transform
$F_{\bbox{\rho}} \equiv \Omega^{-1}{\sum}_{\bf k} f_{\bf k}
\exp (i{\bf k}{\cdot}{\bbox{\rho}})$
of the steady-state distribution takes the form

\begin{equation}
F_{\bbox{\rho}} = { { F_{\bf 0} }\over { F^{\rm eq}_0 } }
 { { F^{\rm eq}_{\rho} } \over
 { 1 - ik_d \rho_x } },
\label{apxT2}
\end{equation}

\noindent
where $k_d = e E\tau/\hbar $
and $F_{\bf 0} = {1\over 2} \langle f \rangle$ per spin state.
Note that,
while a formal distinction is made between $F_{\bf 0}$ and
$F^{\rm eq}_0$, the physical normalisation is always
$F_{\bf 0} = F^{\rm eq}_0 = {1\over 2}{n}$.

The transform of the dynamic BGF,
${\sf {\cal R}}_{\bbox{\rho} \bbox{\rho}' }(\omega) 
\equiv \Omega^{-2} {\sum}_{{\bf k}, {\bf k'}}
{\rm R}_{{\bf k} {\bf k'}}(\omega)
\exp [i({\bf k}{\cdot}{\bbox{\rho}} - {\bf k'}{\cdot}{\bbox{\rho}'})]$,
has the equation

\begin{equation}
\left[
-i\omega\tau - ik_d\rho_x + 1
\right]
{\sf {\cal R}}_{{\bbox{\rho}} {\bbox{\rho}'}}(\omega)
= \tau \delta({\bbox{\rho}} - {\bbox{\rho}'}) +
{ { {\sf {\cal R}}_{{\bf 0} {\bbox{\rho}'}} (\omega) }
  \over
  { F^{\rm eq}_0
} }
F^{\rm eq}_{\rho}.
\label{apxT3}
\end{equation}

\noindent
For ${\bbox{\rho}} = {\bf 0}$ this leads to

\begin{equation}
{\sf {\cal R}}_{{\bf 0} {\bbox{\rho}'}}(\omega)
= -{ {\delta({\bbox{\rho}'}) } \over { i(\omega + i\eta)} }.
\label{apxT4}
\end{equation}

\noindent
On the other hand, the uncorrelated component of 
${\sf {\cal R}}_{{\bbox{\rho}} {\bbox{\rho}'}}$
scales with the steady-state solution $F_{\bbox{\rho}}$
[in a collision-time model the asymptotic form
$F_{\bbox \rho}/{1\over 2}{n}$ replaces
${\Delta F}_{\bbox \rho}/{1\over2}{\Delta {n}}$].
Denoting the correlated part by 
${\sf {\cal C}}_{{\bbox{\rho}} {\bbox{\rho}'}}$
and recalling that the uncorrelated part
exhausts the normalisation of
${\sf {\cal R}}_{{\bf 0} {\bbox{\rho}'}}$,
we obtain

\begin{equation}
{\sf {\cal R}}_{{\bbox{\rho}} {\bbox{\rho}'}}(\omega) =
{\sf {\cal C}}_{{\bbox{\rho}} {\bbox{\rho}'}}(\omega)
- { {\delta({\bbox{\rho}'})}\over {i(\omega + i\eta)} }
{ {F_{\bbox{\rho}}}\over {F_{\bf 0}} }.
\label{apxT5}
\end{equation}

\noindent
When the above is put together with
Eqs. (\ref{apxT2})--(\ref{apxT4}) we arrive,
after some algebra, at the explicit formula
for the correlated propagator:

\begin{equation}
{\sf {\cal C}}_{{\bbox{\rho}} {\bbox{\rho}'}}(\omega) =
\tau { { \delta({\bbox{\rho}} - {\bbox{\rho}'})
     - {\displaystyle { {F_{\bbox{\rho}}}\over
			{F_{\bf 0}} } }
	 \delta({\bbox{\rho}'}) }
 \over
{ 1 - i k_d \rho_x - i \omega \tau } }.
\label{apxT6}
\end{equation}

We can use Eq. (\ref{apxT6}) directly to evaluate both dissipative and
non-dissipative contributions to the noise. Using
the reciprocal-space representation
${\bf v} \leftrightarrow -i(\hbar/m^*)\partial/\partial {\bbox \rho}$,
the power density $P$ of Eq. (\ref{AY16}) is

\begin{eqnarray}
P =&& 2 { {e^2 E^2}\over {k_B T} }
{\left( -{i\hbar\over m^*} \right)}^2
{ \left\{
{ {\partial}\over {\partial \rho_x} }
\int {d^{\nu} \rho'}
{\sf {\cal C}}_{{\bbox \rho} {\bbox \rho'}}(0)
{ {\partial}\over {\partial \rho'_x} }
\Delta F^{\rm eq}_{\rho'}
\right\} }_{\rho \to 0}
\cr
{\left. \right.} \cr
=&& 2 { {e^2 E^2 \tau}\over {k_B T} }
{\left( {\hbar\over m^*} \right)}^2
{ \left\{
-{ {\partial}^2\over {\partial \rho_x^2} }
\Delta F^{\rm eq}_{\rho}
\right\} }_{\rho \to 0}
\cr
{\left. \right.} \cr
=&& \sigma E^2.
\label{apxT8}
\end{eqnarray}

\noindent
The Drude conductivity $\sigma = {n} e \mu_e$
appears when we apply the relation
${\{-\partial^2 \Delta F^{\rm eq}_{\rho}/
\partial \rho_x^2\}}_{\rho \to 0} =
{\langle k^2_x \Delta f^{\rm eq} \rangle} = m^*k_BT{n}/2\hbar^2$
to the middle line of the equation.
A contribution containing
$\langle v_x \Delta f^{\rm eq} \rangle = 0$
vanishes trivially.

The hot-electron spectral density
$S_g$ in the static limit [Eq. (\ref{eq22a})]
is calculated similarly:

\begin{eqnarray}
S_g =&& 2{ {(e{\bf E}{\bbox \cdot}{\bf {\hat x}})^3}\over \hbar}
{ \left\{
\int {d^{\nu} \rho'} \int {d^{\nu} \rho''}
v_x {\sf {\cal C}}_{{\bbox \rho} {\bbox \rho'}}(0)
v'_x {\sf {\cal C}}_{{\bbox \rho'} {\bbox \rho''}}(0)
(-i\rho''_x \Delta F^{\rm eq}_{\rho''})
\right\} }_{\rho \to 0}
\cr
{\left. \right.} \cr
=&& 2{ {e^3 E^3 \tau^2 \hbar}\over {m^*}^2 }
{\left\{
  {\left[
    { {\partial}\over {\partial \rho_x} }
	     { 1\over {1 - ik_d \rho_x} }
    {\left(
      { {\partial}\over {\partial \rho_x} }
      { {-i\rho_x \Delta F^{\rm eq}_{\rho}}\over
	{1 - ik_d \rho_x} }
    \right)}
  \right]}_{\rho \to 0}
\right.}
\cr
&&{\left. - {\left[
{ {\partial}\over {\partial \rho_x} }
{ {F_{\bbox \rho}/F_{\bf 0}}\over
  {1 - ik_d \rho_x} } \right]}_{\rho \to 0}
   {\left[
{ {\partial}\over {\partial \rho'_x} }
{ {-i\rho'_x \Delta F^{\rm eq}_{\rho'}}\over
   {1 - ik_d \rho'_x} } \right]}_{\rho' \to 0}
\right\} }.
\label{apxT9}
\end{eqnarray}

\noindent
We evaluate this with the help of the relations
$\Delta F^{\rm eq}_0 = {1\over 2}\Delta {n}$ and
$\{ \partial F_{\bbox \rho}/\partial \rho_x \}_{\rho \to 0}
= ik_dF_{\bf 0}$, the latter following from Eq. (\ref{apxT2}).
The result is

\begin{eqnarray}
S_g
=&& \sigma m^* \mu_e^2 E^4 {{\Delta {n}}\over {n}}.
\label{apxT10}
\end{eqnarray}

\section{Weakly Nonuniform Drude Model}
\label{apxSN}

We derive the spatio-temporal correlations within
the Drude model of the preceding Appendix.
The problem is to calculate the
propagation of a single electron added to $N$ uniformly
distributed electrons at a {\em specific}
point in the sample at $t=0$. This constitutes a weak inhomogeneity.

Since the scattering is spin-independent, we consider a
zero-spin model with effective density ${1\over 2}n$.
The equation of motion for the dynamical propagator
in the frequency domain is

\begin{eqnarray}
{\left[ 
\tau {\bf v}_{\bf k} {\bbox \cdot}
{ {\partial}\over {\partial {\bf r}} }
+ k_d { {\partial}\over {\partial k_x} }
+ 1 - i\omega \tau
\right]}
{\rm R}_{\alpha \alpha'}(\omega)
=&& \tau \delta_{{\bf r} {\bf r'}} \delta_{{\bf k} {\bf k'}}
+  { {\sum_{\alpha''}{\rm R}_{\alpha'' \alpha'}(\omega)}
\over {{1\over 2}N} } f^{\rm eq}_k;
\label{apxSN1}
\end{eqnarray}

\noindent
Now define the Fourier transform of the propagator,

\begin{equation}
{\rm R}_{{\bf k} {\bf k'}}({\bf q}, {\bf q'}, \omega)
=
\int_{\Omega} \!\! d^{\nu} r \int_{\Omega} \!\! d^{\nu} r' {~}
{\rm R}_{\alpha \alpha'}(\omega)
\exp [-i({\bf q}{\bbox \cdot}{\bf r}
         - {\bf q'}{\bbox \cdot}{\bf r'})].
\label{apxSN2}
\end{equation}

\noindent
Eq. (\ref{apxSN1}) becomes

\begin{eqnarray}
{\left[ 
k_d { {\partial}\over {\partial k_x} }
+ { {i\hbar \tau}\over m^* } {\bf q} {\bbox \cdot} {\bf k}
+ 1 - i\omega \tau
\right]}
{\rm R}_{{\bf k} {\bf k'}}({\bf q}, {\bf q'}, \omega)
=&&
\tau \Omega \delta_{{\bf k} {\bf k'}}
\Omega \varphi_{\nu}({\bf q} - {\bf q'})
\cr
&&
+ {\langle {\rm R}({\bf 0}, {\bf q'}, \omega; {\bf k'}) \rangle}
\Omega \varphi_{\nu}({\bf q}) { {f^{\rm eq}_k}\over {{1\over 2}N} },
\label{apxSN3}
\end{eqnarray}

\noindent
where

\begin{equation}
\varphi_{\nu}({\bf q})
= {1\over \Omega} \int_{\Omega} d^{\nu}r
e^{-i{\bf q}{\bbox \cdot}{\bf r}}.
\label{apxSNf1}
\end{equation}

\noindent
For a three-dimensional system
with cylindrical symmetry about the {\em x}-axis we write
${\bf q} = (q_x, {\bf q}_{\perp})$
and $\Omega = \pi R^2 L_x$ where $L_x$ is the sample length
and $R$ its radius. The function $\varphi_3$ can be written as

\begin{equation}
\varphi_3({\bf q})
= \varphi_1(q_x; {\textstyle {1\over 2}} L_x)
\varphi_2({\bf q}_{\perp}; R)
\equiv { {\sin ({1\over 2}L_x q_x)}\over {{1\over 2}L_x q_x} }
{ J_1(Rq_{\perp})\over {{1\over 2}R q_{\perp}} }
\label{apxSNf2}
\end{equation}

\noindent
with $J_1(u)$ the first-order Bessel function.
For a two-dimensional strip of width $2R$, the role of
$\varphi_3$ is assumed by the product
$\varphi_1(q_x; {1\over 2}L_x) \varphi_1(q_{\perp}; R)$.

The correlated dynamical propagator ${\rm C}$ associated
with ${\rm R}$ has the form

\begin{eqnarray}
{\rm C}_{{\bf k} {\bf k'}}({\bf q}, {\bf q'}, \omega)
=&& {\rm R}_{{\bf k} {\bf k'}}({\bf q}, {\bf q'}, \omega)
- {\langle {\rm R}({\bf 0}, {\bf q'}, \omega; {\bf k'}) \rangle}
\varphi_{\nu}({\bf q}){f_{\bf k}\over {{1\over 2}n}}
\cr
{\left. \right.} \cr
=&& {\rm R}_{{\bf k} {\bf k'}}({\bf q}, {\bf q'}, \omega)
+ { {\Omega \varphi_{\nu}({\bf q}) \varphi_{\nu}({\bf q'})}\over
    {i(\omega + i\eta)} }
{f_{\bf k}\over {{1\over 2}n}},
\label{apxSN4}
\end{eqnarray}

\noindent
wherein ${\langle {\rm R} \rangle}$ is evaluated
by summing over ${\bf k}$ on both sides of Eq. (\ref{apxSN3})
in the limit $q \to 0$. The equation of motion for ${\rm C}$ is
\cite{kogan}

\begin{eqnarray}
{\Biggl[
\Biggr.}&&{\Biggl. 
k_d { {\partial}\over {\partial k_x} }
+ { {i\hbar \tau}\over m^* } {\bf q} {\bbox \cdot} {\bf k}
+ 1 - i\omega \tau
\Biggr]}
{\rm C}_{{\bf k} {\bf k'}}({\bf q}, {\bf q'}, \omega)
\cr
{\left. \right.} \cr
&&=
\tau \Omega^2 \delta_{{\bf k} {\bf k'}}
\varphi_{\nu}({\bf q} - {\bf q'})
+ {\langle {\rm R} ({\bf 0}, {\bf q'}, \omega; {\bf k'}) \rangle}
\varphi_{\nu}({\bf q}) {f^{\rm eq}_k\over {{1\over 2}{n}} }
\cr
&& {~~~}
- {\langle {\rm R} ({\bf 0}, {\bf q'}, \omega; {\bf k'}) \rangle}
\varphi_{\nu}({\bf q})
{\left[
k_d { {\partial}\over {\partial k_x} }
+ 1 - i\omega\tau
\right]}
{ {f_{\bf k}}\over {{1\over 2}{n}} }
{\left. \right.} \cr
\cr
&&=
\tau \Omega {\left( \Omega \delta_{{\bf k} {\bf k'}}
\varphi_{\nu}({\bf q} - {\bf q'})
- \varphi_{\nu}({\bf q}) \varphi_{\nu}({\bf q'})
{f_{\bf k}\over {{1\over 2}{n}}}
\right)}.
\label{apxSN5}
\end{eqnarray}

\noindent
Terms $ \sim 1/\omega$, singular in the static limit,
cancel identically by the fact that $f$ is the Boltzmann solution
for the uniform steady state.
Equation (\ref{apxSN5}) is solved at arbitrary fields
with the integrating factor
\cite{sw0}

\begin{equation}
X_{\bf k} ({\bf q}, \omega) =
\exp \!{\left\{ {k_x\over k_d}
      {\left[
{{i\hbar\tau}\over m^*}
{\left(
{{q_x k_x}\over 2} + {\bf q}_{\perp} {\bbox \cdot}{\bf k}_{\perp}
\right)}
+ 1 - i\omega \tau
       \right]}
     \right\} };
\label{apxSN6}
\end{equation}

\noindent
in the low-field limit, the non-analytic character of $X$
is clear from the occurrence of $1/k_d \propto 1/E$
in its exponent. Using $X$ we first generate
the auxiliary propagator ${\rm C}^{(0)}$ satisfying

\begin{mathletters}
\label{apxSN6.1}
\begin{equation}
{\Biggl[
k_d { {\partial}\over {\partial k_x} }
+ { {i\hbar \tau}\over m^* } {\bf q} {\bbox \cdot} {\bf k}
+ 1 - i\omega \tau
\Biggr]}
{\rm C}^{(0)}_{{\bf k} {\bf k'}}({\bf q}, \omega)
= \tau \Omega \delta_{{\bf k} {\bf k'}}.
\label{apxSN6.1a}
\end{equation}

\noindent
The expression for ${\rm C}^{(0)}$ is

\begin{equation}
{\rm C}^{(0)}_{{\bf k} {\bf k'}}({\bf q}, \omega)
= {{\tau \Omega}\over {k_d L_x}}
\delta_{{\bf k}_{\perp} {\bf k}'_{\perp}}
\theta(k_x - k'_x) X^{-1}_{\bf k}({\bf q}, \omega)
X_{\bf k'}({\bf q}, \omega),
\label{apxSN6.1b}
\end{equation}
\end{mathletters}

\noindent
which furnishes the complete solution to Eq. (\ref{apxSN5}) as

\begin{equation}
{\rm C}_{{\bf k} {\bf k'}} ({\bf q}, {\bf q'}, \omega)
= \Omega \varphi_{\nu}({\bf q} - {\bf q'})
{\rm C}^{(0)}_{{\bf k} {\bf k'}} ({\bf q}, \omega)
- {{\varphi_{\nu}({\bf q}) \varphi_{\nu}({\bf q'})}\over
  {{1\over 2}n}}
\sum_{\bf k''}
{\rm C}^{(0)}_{{\bf k} {\bf k''}} ({\bf q}, \omega)
f_{\bf k''}.
\label{apxSN7}
\end{equation}

For our study of shot noise it is convenient to consider
the propagator integrated over ${\bf q'}$, namely

\begin{equation}
{\rm C}^{(b)}_{{\bf k} {\bf k'}} ({\bf q}, \omega)
= \int {{d^{\nu} q'}\over (2\pi)^{\nu}}
{\rm C}_{{\bf k} {\bf k'}}({\bf q}, {\bf q'}, \omega),
\label{apxSN5.0}
\end{equation}

\noindent
equivalent to the bulk solution in an infinitely wide conductor,
$R \to \infty$.
The form of ${\rm C}^{(b)}$ follows from Eq. (\ref{apxSN7}). It is

\begin{equation}
{\rm C}^{(b)}_{{\bf k} {\bf k'}} ({\bf q}, \omega)
= {\rm C}^{(0)}_{{\bf k} {\bf k'}} ({\bf q}, \omega)
- {{\varphi_{\nu}({\bf q})}\over {{1\over 2}n\Omega}}
\sum_{\bf k''}
{\rm C}^{(0)}_{{\bf k} {\bf k''}} ({\bf q}, \omega)
f_{\bf k''}.
\label{apxSN7.0}
\end{equation}

\noindent
In the zero-field limit this goes to

\begin{equation}
{\rm C}^{(b)}_{{\bf k} {\bf k'}} ({\bf q}, \omega)\Big|_{E \to 0}
= { { \Omega \delta_{{\bf k} {\bf k'}}
      - \varphi_{\nu}({\bf q})
	{\displaystyle {f^{\rm eq}_k\over {1\over 2}{n}}} }
	\over
	{ {\displaystyle {{i\hbar}\over m^*} }
	{\bf q}{\bbox \cdot}{\bf k}
	+ \tau^{-1} - i\omega } }.
\label{apxSN8}
\end{equation}

Finally, for reference, we also record the
expression for the uniform distribution $\Delta f_{\bf k}$,
needed in the shot-noise application:

\begin{mathletters}
\label{apxSN13}
\begin{equation}
\Delta f_{\bf k}
= \int^{k_x}_{-\infty} {dk'_x\over k_d} e^{-(k_x - k'_x)/k_d}
\Delta f^{\rm eq}_{k'}.
\label{apxSN13a}
\end{equation}

\noindent
In the degenerate limit this becomes

\begin{equation}
\Delta f_{\bf k}
= {{m^* k_BT}\over {\hbar^2 p_{\perp} k_d}}
\theta(k_F - k_{\perp})
{\left[
  \theta(k_x - p_{\perp}) e^{(p_{\perp} - k_x)/k_d}
+ \theta(k_x + p_{\perp}) e^{-(p_{\perp} + k_x)/k_d}
\right]},
\label{apxSN13b}
\end{equation}
\end{mathletters}

\noindent
where $p_{\perp} = (k^2_F - k^2_{\perp})^{1\over 2}$.
In one dimension, set $k_{\perp} = 0$.

\newpage

\end{document}